\documentclass[default,icol]{sn-jnl}

\jyear{2023}%

\theoremstyle{thmstyleone}%
\theoremstyle{thmstyletwo}%

\theoremstyle{thmstylethree}%

\begin{document}

\title[Niel's Chess]{Niel's Chess\texttrademark\,\,- Rules for Xiangqi}

\author*{\fnm{Tamás} \sur{Varga}}\email{tvarga@q-edu-lab.com}
\affil{\orgname{q-edu-lab.com}, \orgaddress{\state{Zurich}, \country{Switzerland}}}


\maketitle


\section{Background}\label{intro}

In this paper, the rules of Niel's Chess \cite{varga} are adapted to the game of Xiangqi, following the idea that the River and the Palaces play an important role in restricting and enabling chess pieces in their movements.

\section{Playing the game}\label{playing}

The "Playing the game" chapter of the World Xiangqi Rules document \cite{wxf} is taken as baseline, in which Articles 1 to 4 describe the non-competitive rules of Xiangqi. Those articles are extended here by the additional Articles 5 to 7, specifying the non-competitive quantum rules of the Niel's Chess variant of Xiangqi. The addition of quantum rules enriches the game with creative ideas in both attack and defence.
\begin{figure}[H]
\centering
\includegraphics[width=0.45\columnwidth]{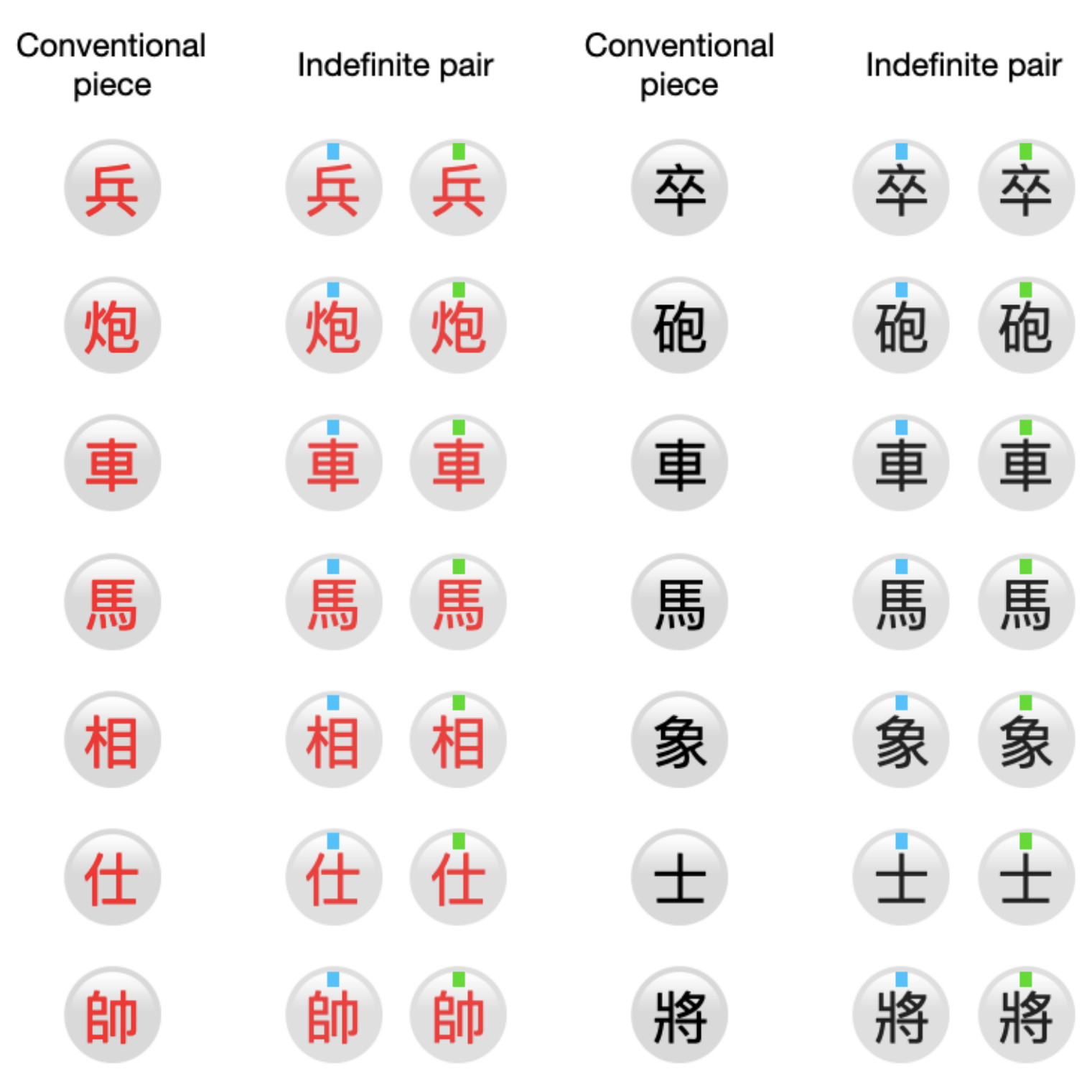}
\end{figure}

\subsection*{Article 5: Indefinite pieces}

\begin{enumerate}[1.1]
\item[5.1] For every conventional piece there is a corresponding pair of 'indefinite pieces', including one piece with a blue mark on it and another with green.\\
\item[5.2] Articles 1 to 4 apply to situations involving indefinite pieces, unless specified otherwise in Articles 6 and 7. Articles 5.2.1 to 5.2.4 stand here only to help understanding:\\
\begin{enumerate}[1.1.1]
\item[5.2.1] \textit{[Clarification]} In accordance with Articles 2.8 to 2.10, 3.1.A.III and 4.5, leaving one’s own indefinite King under threat, exposing one’s own indefinite King to threat, capturing the opponent’s indefinite King, and also making one's own indefinite King face the opponent's King, indefinite or not, on the same file without any intervening piece, is not allowed, barring Articles 7.9 and 7.10, unless it falls under the concept 'committing suicide'.
\item[5.2.2] \textit{[Clarification]} An indefinite piece may move to an intersection in accordance with the rules for conventional pieces stipulated in Articles 2.1 to 2.9, with the exception that the capturing part of such a move, if any, is governed by Articles 6 and 7. It's not allowed to move over blocking indefinite pieces, and indefinite pieces may be used as Cannon mounts.
\item[5.2.3] \textit{[Clarification]} A piece, indefinite or not, is said to threaten an opponent’s piece, indefinite or not, if the piece could make (or at least attempt, see Article 6.4) a capture on that intersection according to Articles 2.1 to 2.8 (disregarding any other rule, but noting Articles 7.9 and 7.10).
\item[5.2.4] \textit{[Clarification]} In accordance with Article 1.4, initially there are no indefinite pieces on the chessboard.
\begin{figure}[H]
\centering
\includegraphics[width=0.45\columnwidth]{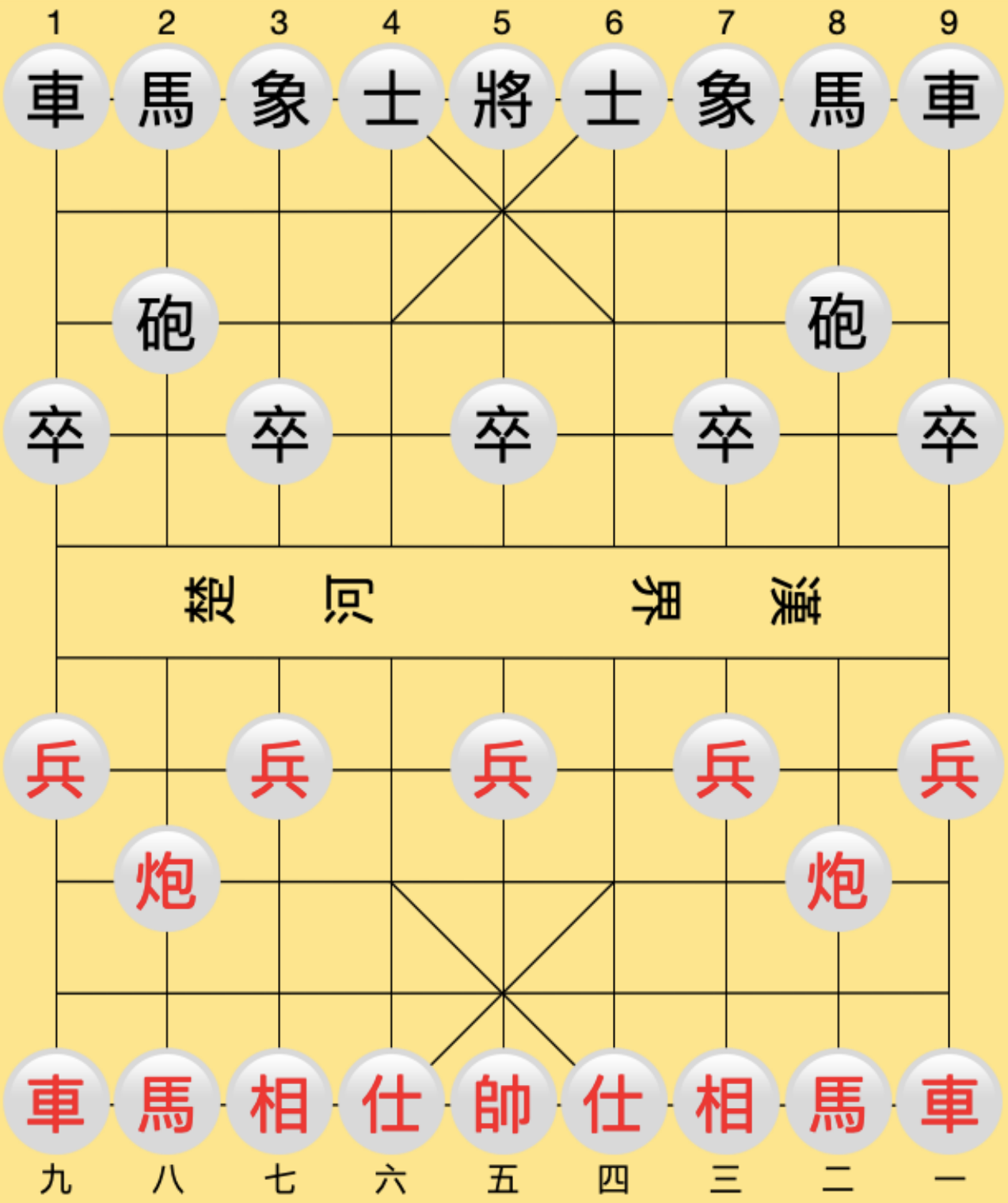}
\end{figure}
\end{enumerate}
\item[5.3] All instances of indefinite pairs of the same piece and colour must have their marks differentiated from those of the others, e.g. by labelling. (This is to avoid ambiguity as to which indefinite pieces on the board are paired, see Article 6.1.)\\
\begin{enumerate}[1.1.1]
\item[5.3.1] \textit{[Recommendation]} For Pawn pair instances, it is recommended to use the numbers 1, 3, 5, 7 and 9 as labels on the blue and green marks. For the rest, there can be at most two pair instances on the board, so a simple dot would suffice to differentiate them. See the "close-up" figures below.
\end{enumerate}
\begin{figure}[H]
\centering
\includegraphics[width=0.45\columnwidth]{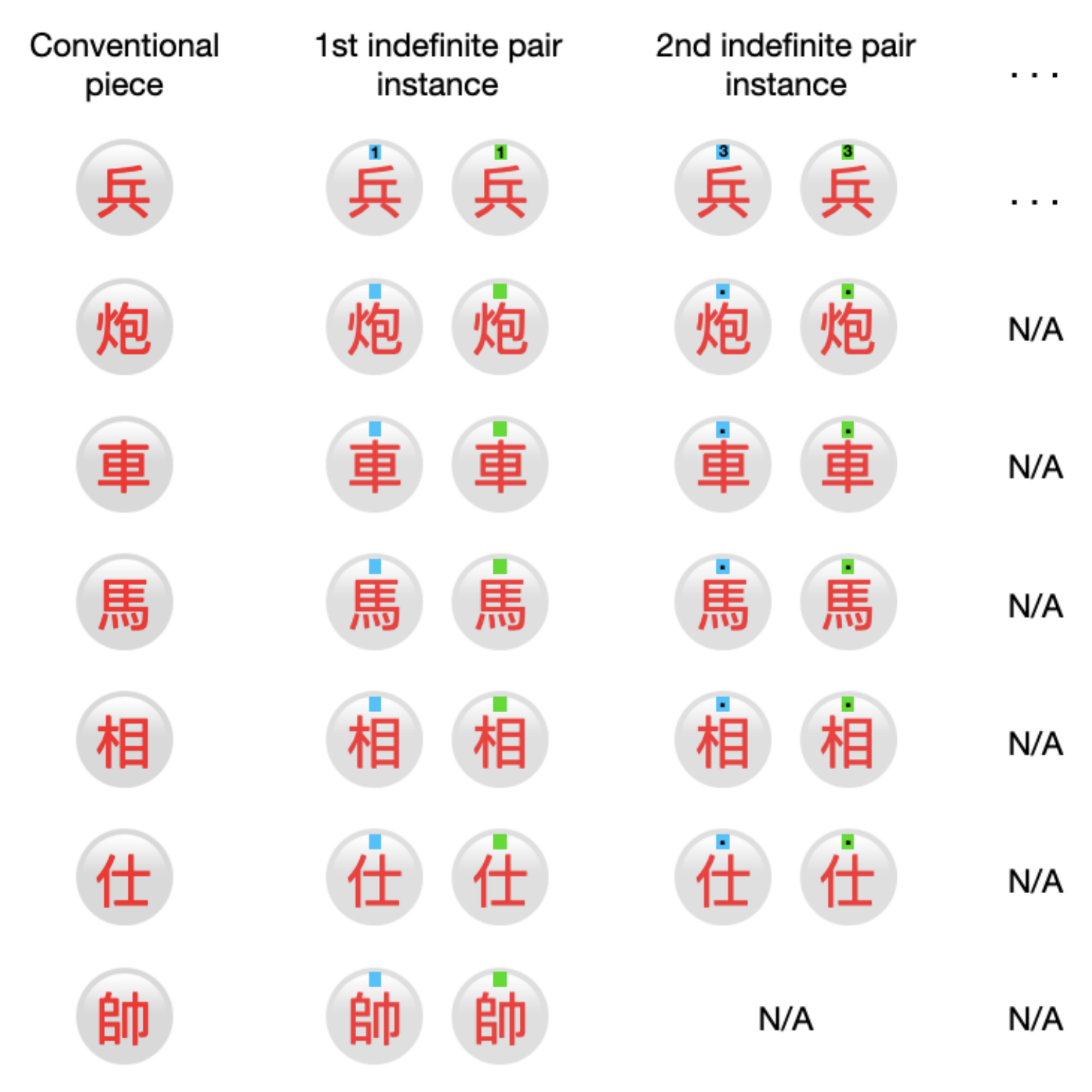}\hfill\includegraphics[width=0.45\columnwidth]{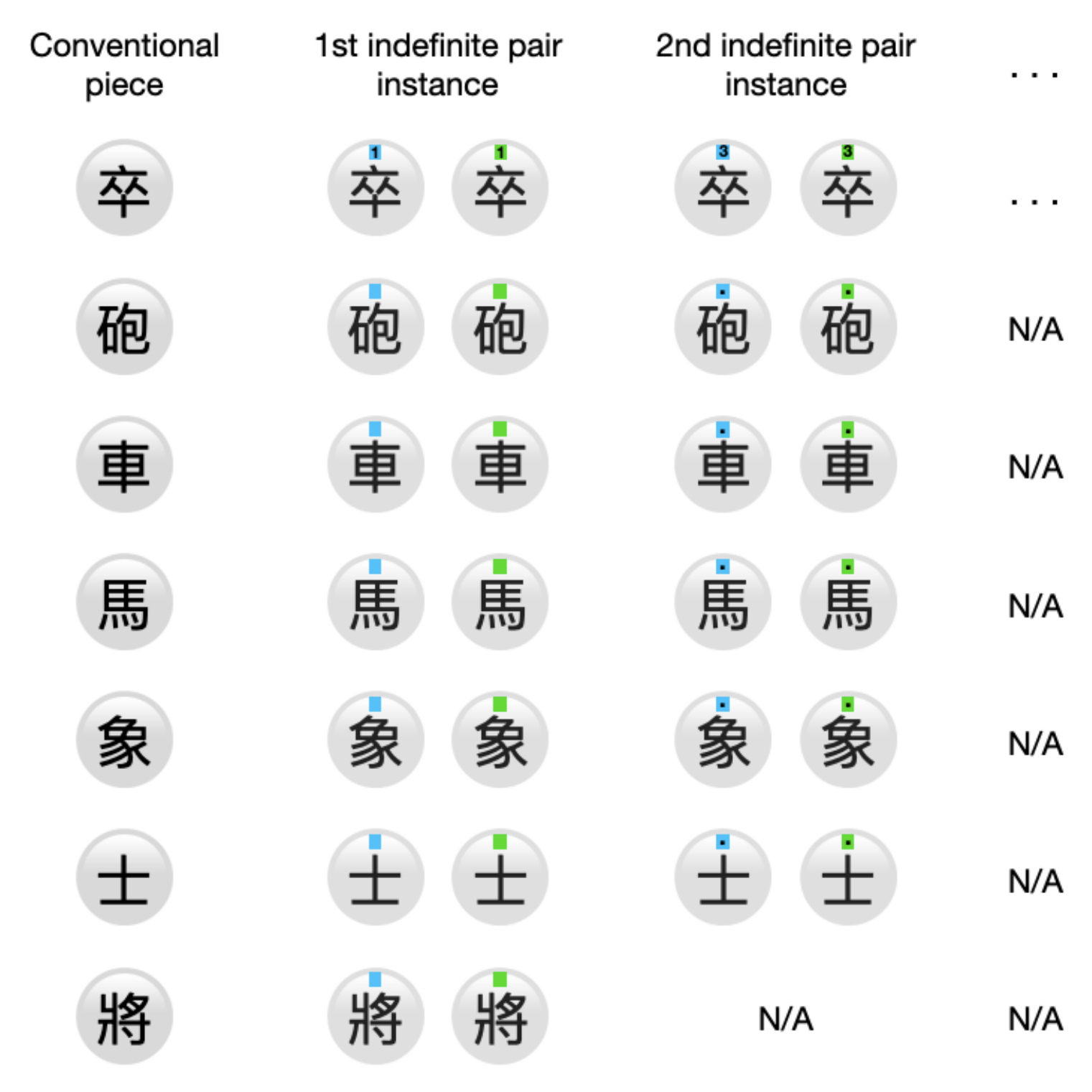}
\vspace*{-0.75cm}
\end{figure}
\begin{figure}[H]
\centering
\includegraphics[width=0.45\columnwidth]{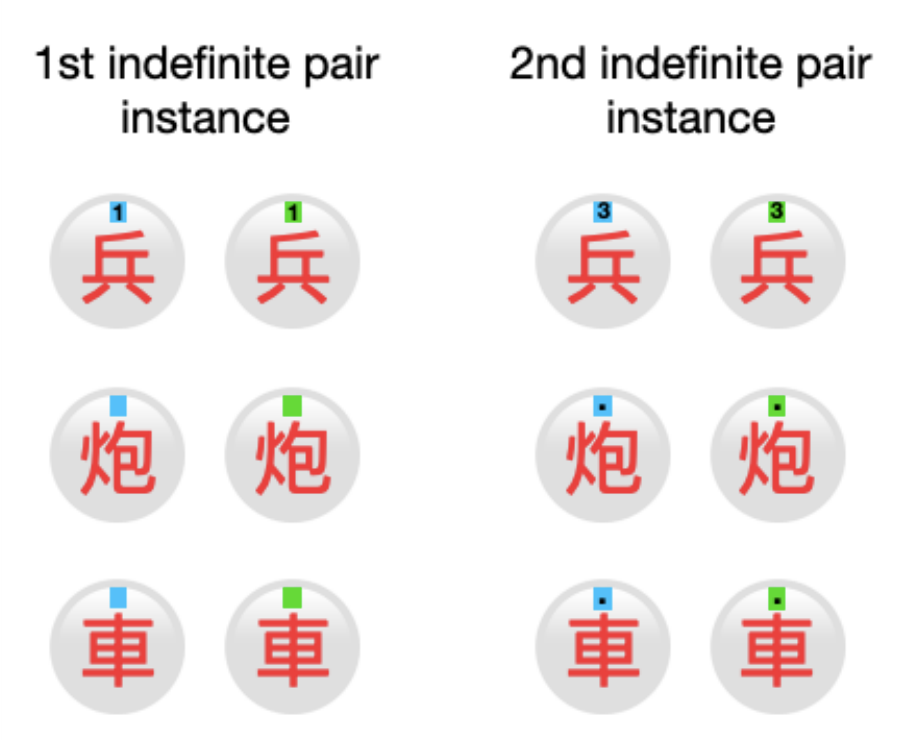}\hfill\includegraphics[width=0.45\columnwidth]{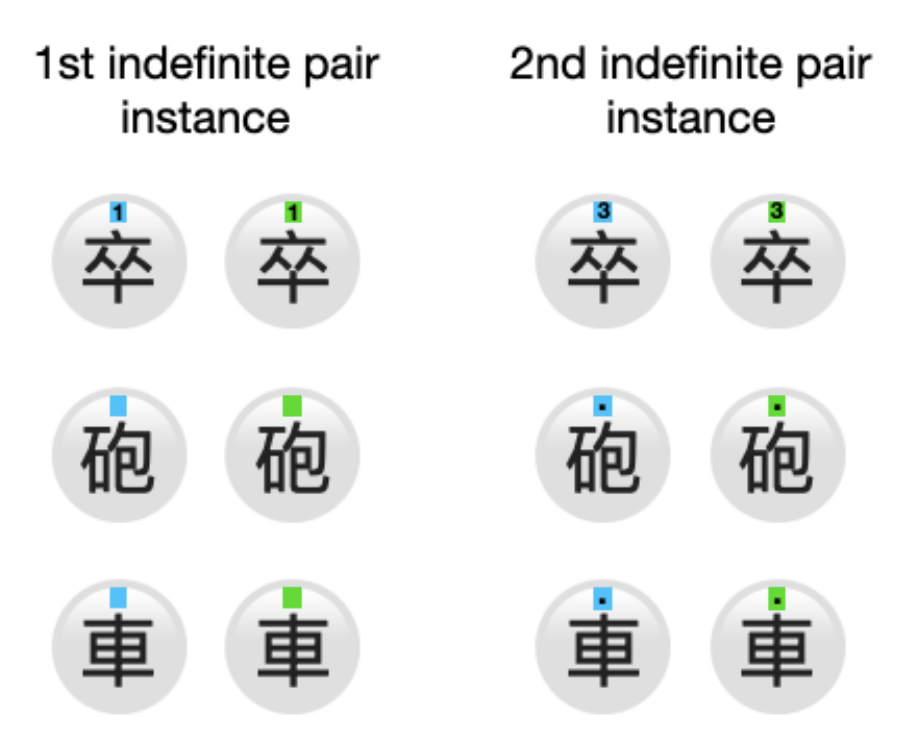}
\end{figure}
\end{enumerate}

\subsection*{Article 6: Superposition rules}

\begin{enumerate}[1.1]
\item[6.1] A conventional piece may move to two unoccupied intersections at once (each reachable as per Articles 2.1 to 2.9, but see also Article 6.3), or move to one unoccupied intersection and stay where it is at once. This is called a 'superposition move'. The conventional piece has to be replaced by the two pieces of a corresponding indefinite pair instance, placed on the two intersections of arrival, one each. The two indefinite pieces are said to be 'paired' (as they represent the conventional piece in a superposition state of being on two intersections at once).
\begin{figure}[H]
\centering
\includegraphics[width=0.45\columnwidth]{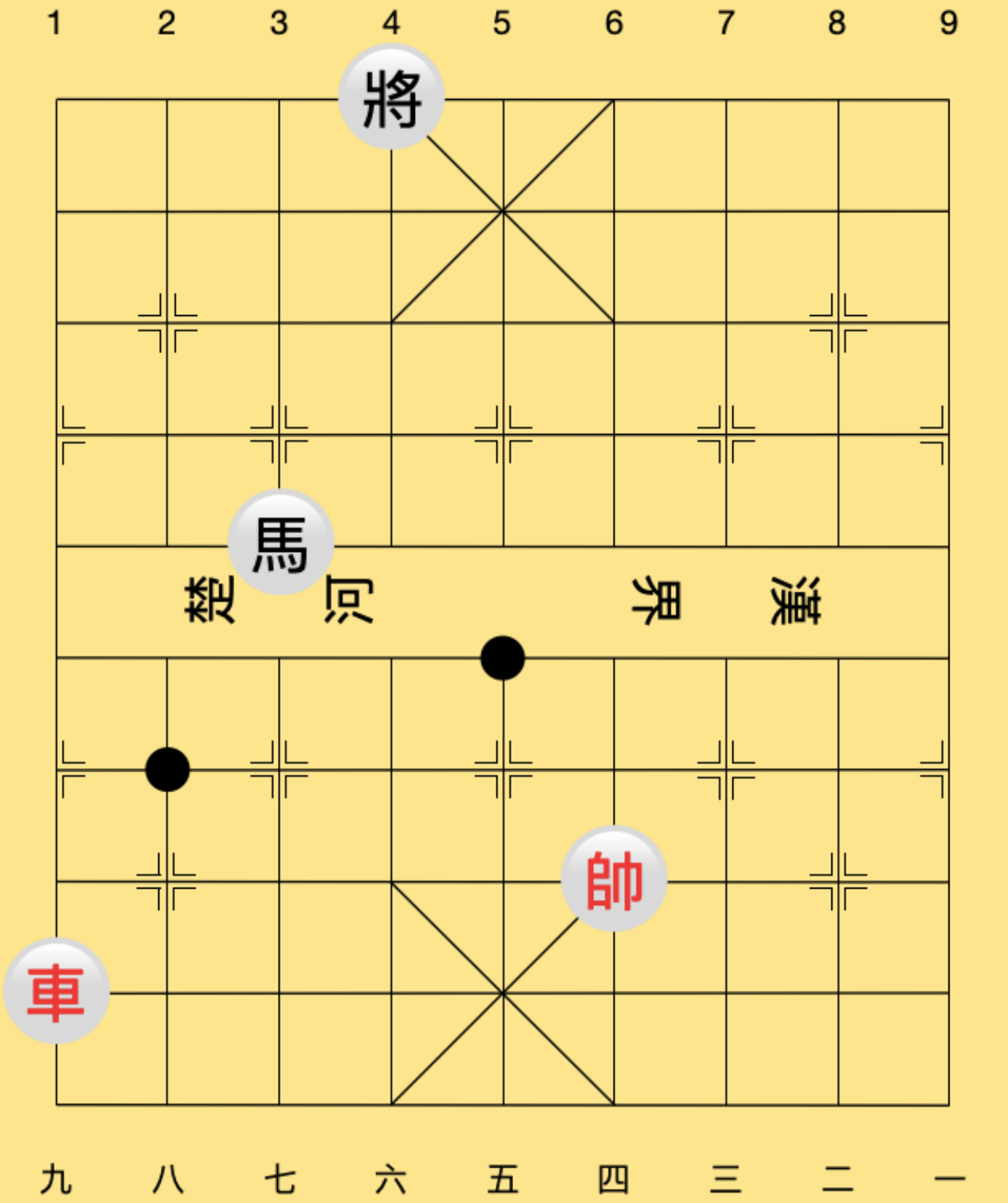}
\end{figure}
\item[6.2] If the marks on the paired indefinite pieces both face the opponent, or both face the player who owns the pair, it is said to be an 'equal superposition' (using opponent-facing marks are recommended for an easier overview\footnote{In this paper, we'll prefer player-facing marks for Black, due to better readability.}). If one mark faces the opponent and the other the player who owns the pair, it is called an 'unequal superposition'. The player making the superposition move can freely choose either option. No other possibilities are allowed (i.e. only facing the upper or lower side of the board is permitted).
\begin{figure}[H]
\centering
\includegraphics[width=0.45\columnwidth]{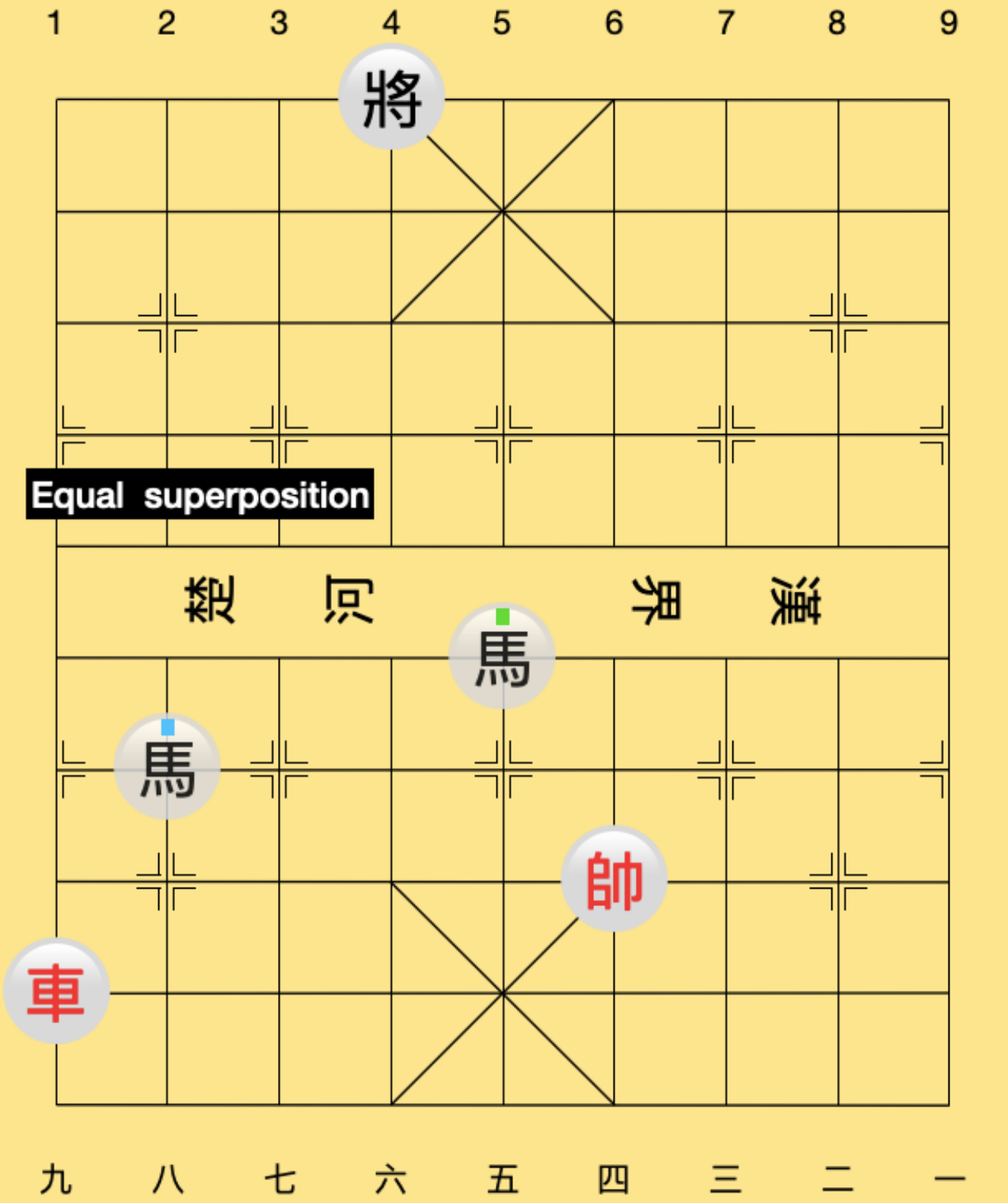}\hfill\includegraphics[width=0.45\columnwidth]{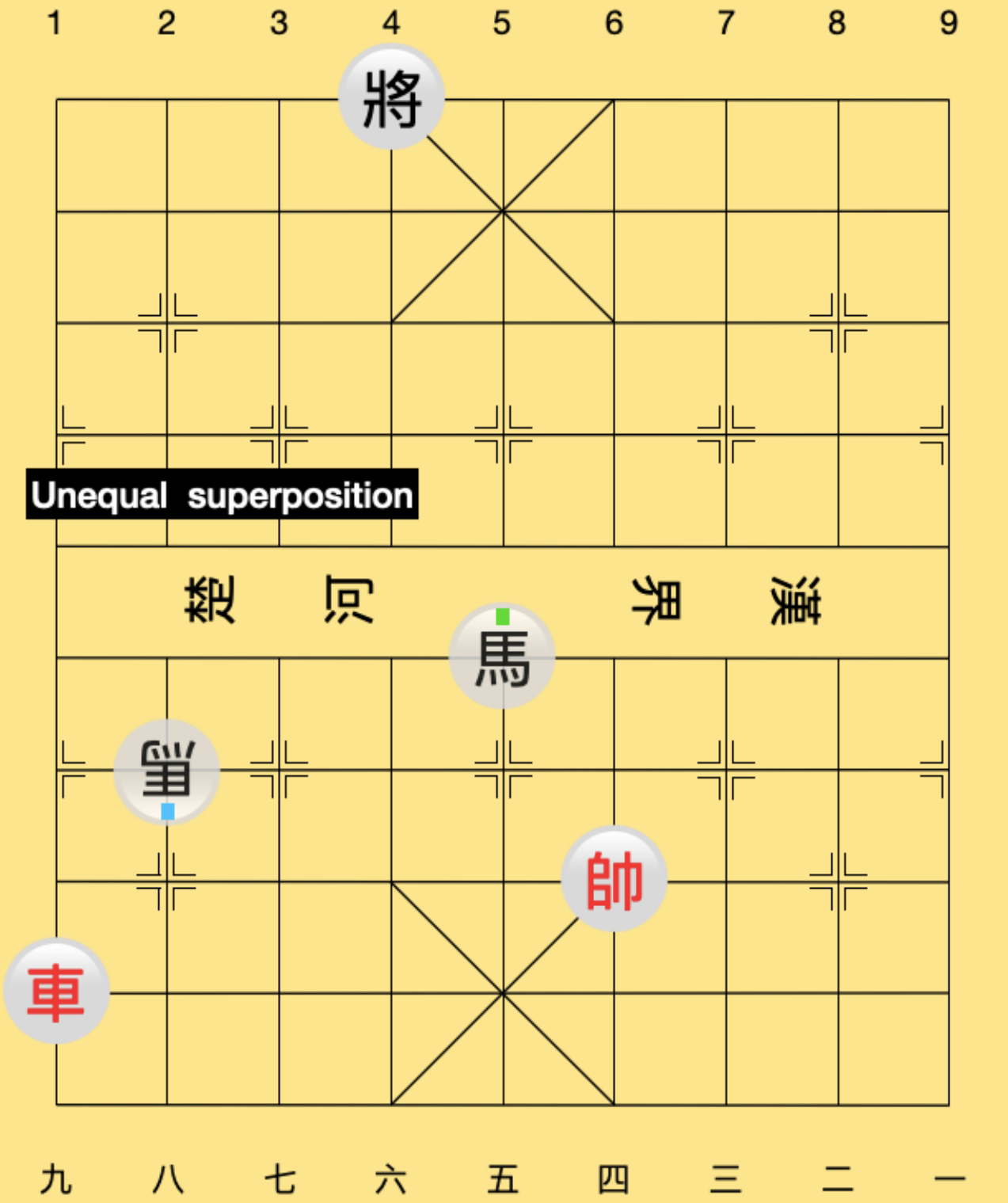}
\end{figure}
\item[6.3] For Horse, Cannon and Chariot, a superposition move is allowed only if at least one intersection of arrival is on the opposite side of the River from the departure. For Adviser, Elephant and King, the precondition for a superposition move is that the intersection of departure belongs to the Palace. For Pawn, it may make a superposition move only after it has crossed the River.
\begin{figure}[H]
\centering
\includegraphics[width=0.45\columnwidth]{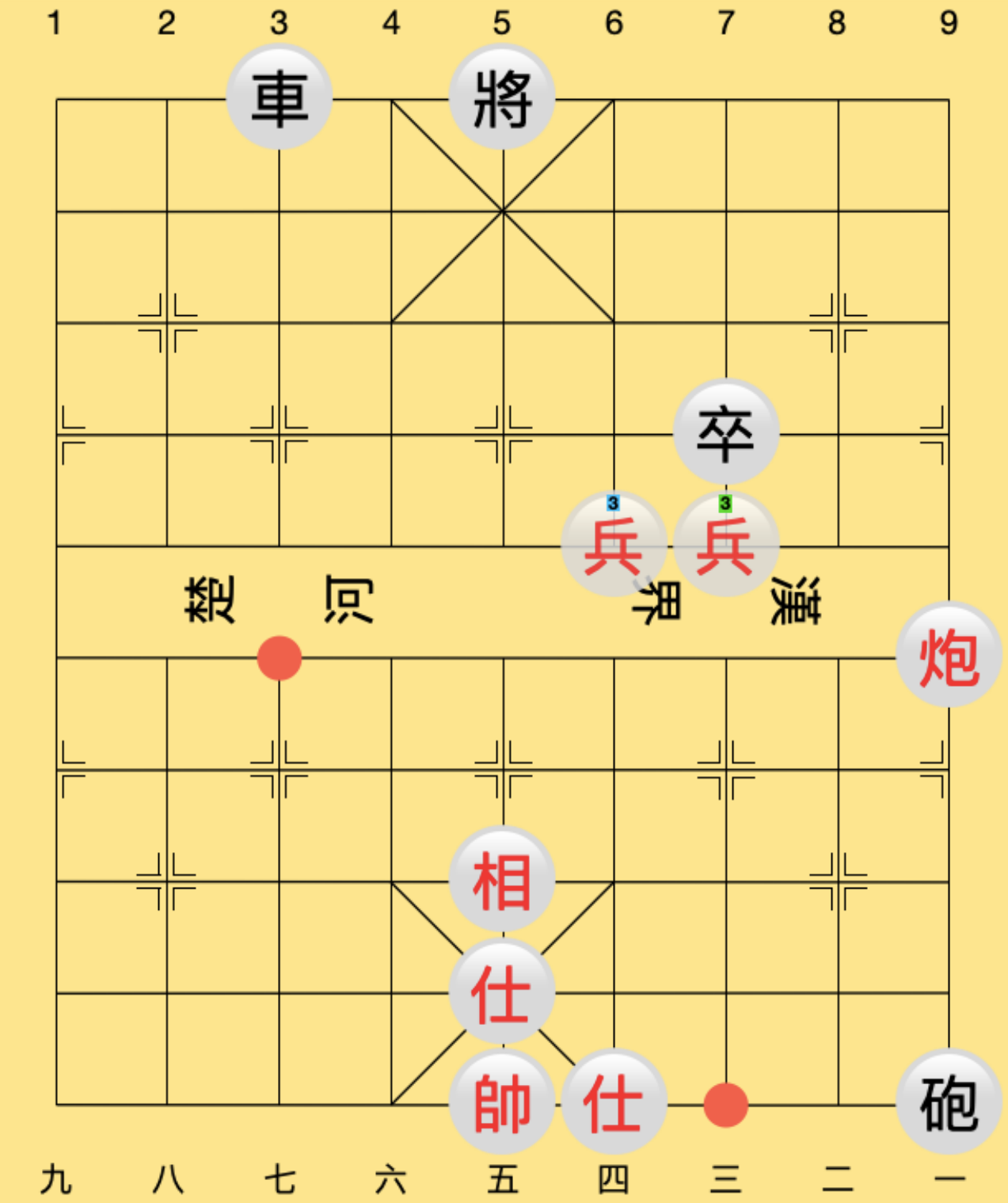}
\end{figure}
\item[6.4] A capture in which at least one indefinite pair instance is involved is called an 'attempted capture'.\\
\begin{enumerate}[1.1.1]
\item[6.4.1] \textit{[Clarification]} An indefinite piece may attempt to capture not only another indefinite piece but also a conventional piece, and vice versa.
\end{enumerate}
\begin{figure}[H]
\centering
\includegraphics[width=0.45\columnwidth]{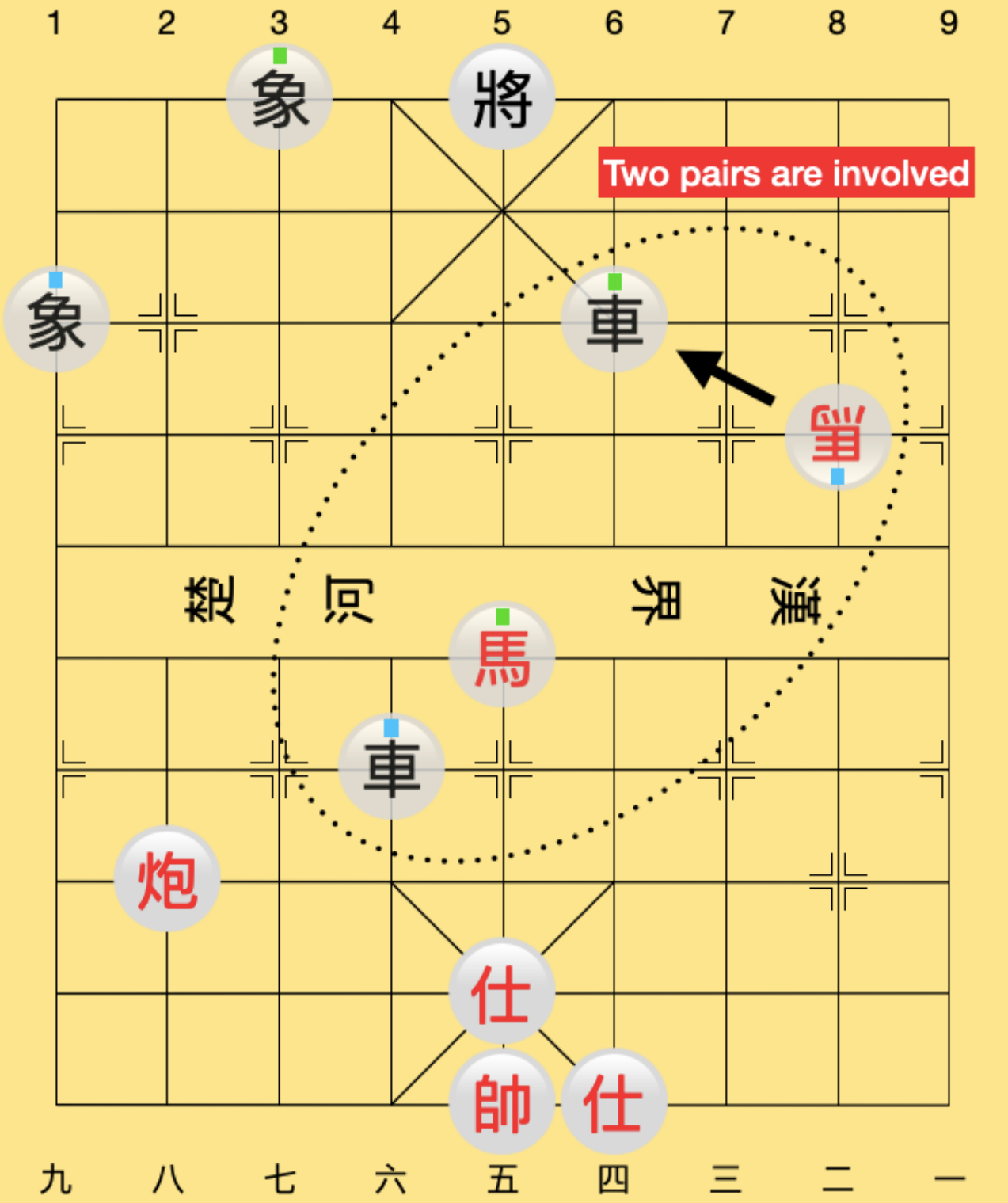}
\end{figure}
\item[6.5] Steps to execute the attempted capture (see also Article 7.5):\\
\begin{enumerate}[1.]
\item[1)] The player attempting the capture places his/her piece on the intended intersection of arrival. (Next to the opponent's piece, there may be overlap if needed.)
\begin{figure}[H]
\centering
\includegraphics[width=0.45\columnwidth]{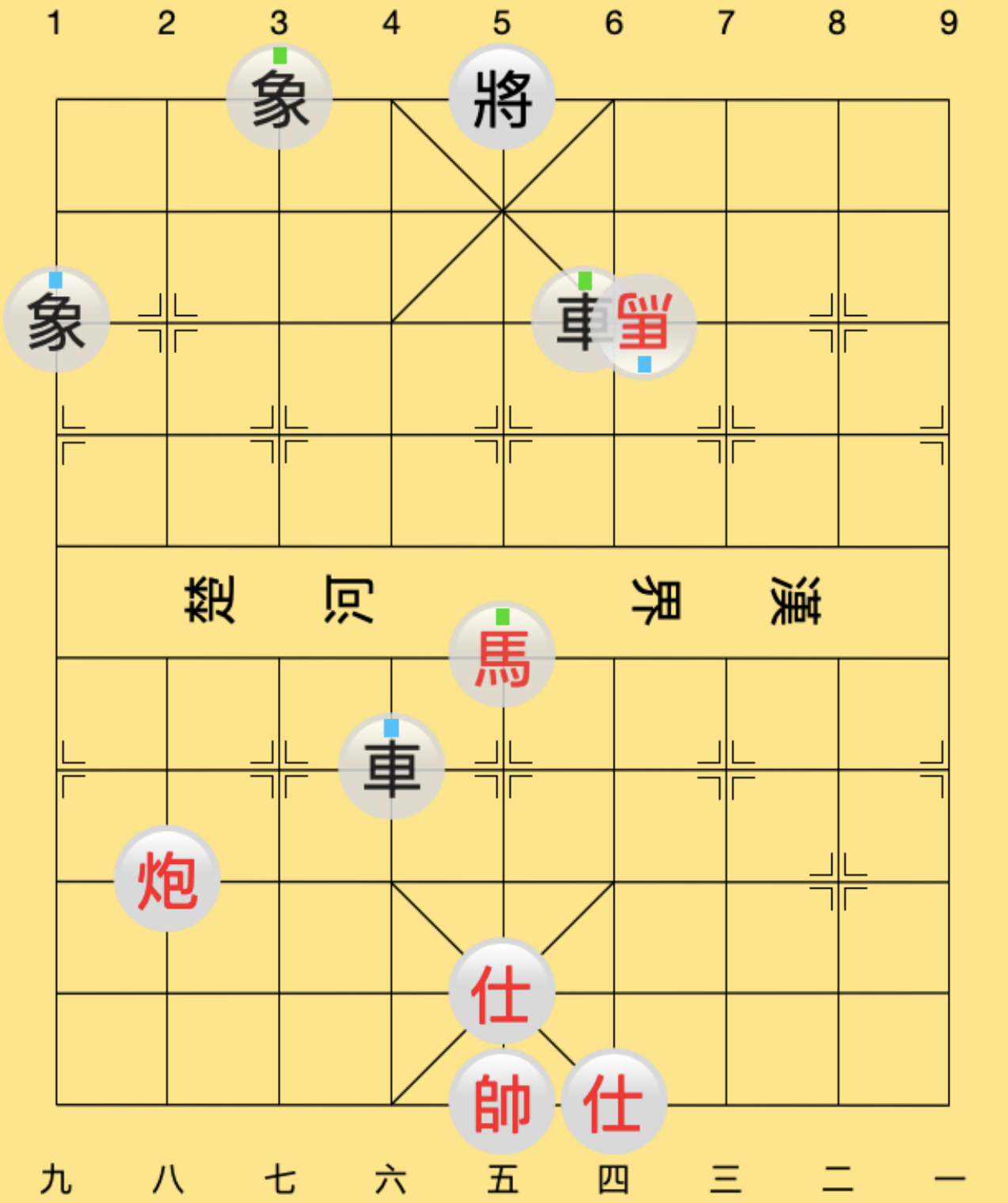}
\end{figure}
\item[2)] Each player who has an indefinite piece on the intended intersection of arrival rolls a six-sided dice:\\\\
- \textbf{Equal superposition with opponent-facing marks: }an even result 2, 4 or 6 (an odd result 1, 3 or 5) means that the player's indefinite pair instance 'collapses' to the intersection having the indefinite piece with the blue mark (the green mark) on it.\\\\
- \textbf{Equal superposition with player-facing marks: }same instructions as for the opponent-facing case, except that the words "blue" and "green" are swapped.\\\\
- \textbf{Unequal superposition: }a result 1, 2, 3 or 4 (a result 5 or 6) means that the player's indefinite pair instance 'collapses' to the intersection having the indefinite piece on it whose mark faces the opponent (the player).\\\\
- The two indefinite pieces are \textbf{replaced by the corresponding conventional piece} that must be placed on the intersection to which the pair has collapsed (as collapse means that the conventional piece has a definite position now).\\\\
- \textbf{Order: }if both players have to roll the dice, the one whose piece is attempted to be captured goes first.
\begin{figure}[H]
\centering
\includegraphics[width=0.45\columnwidth]{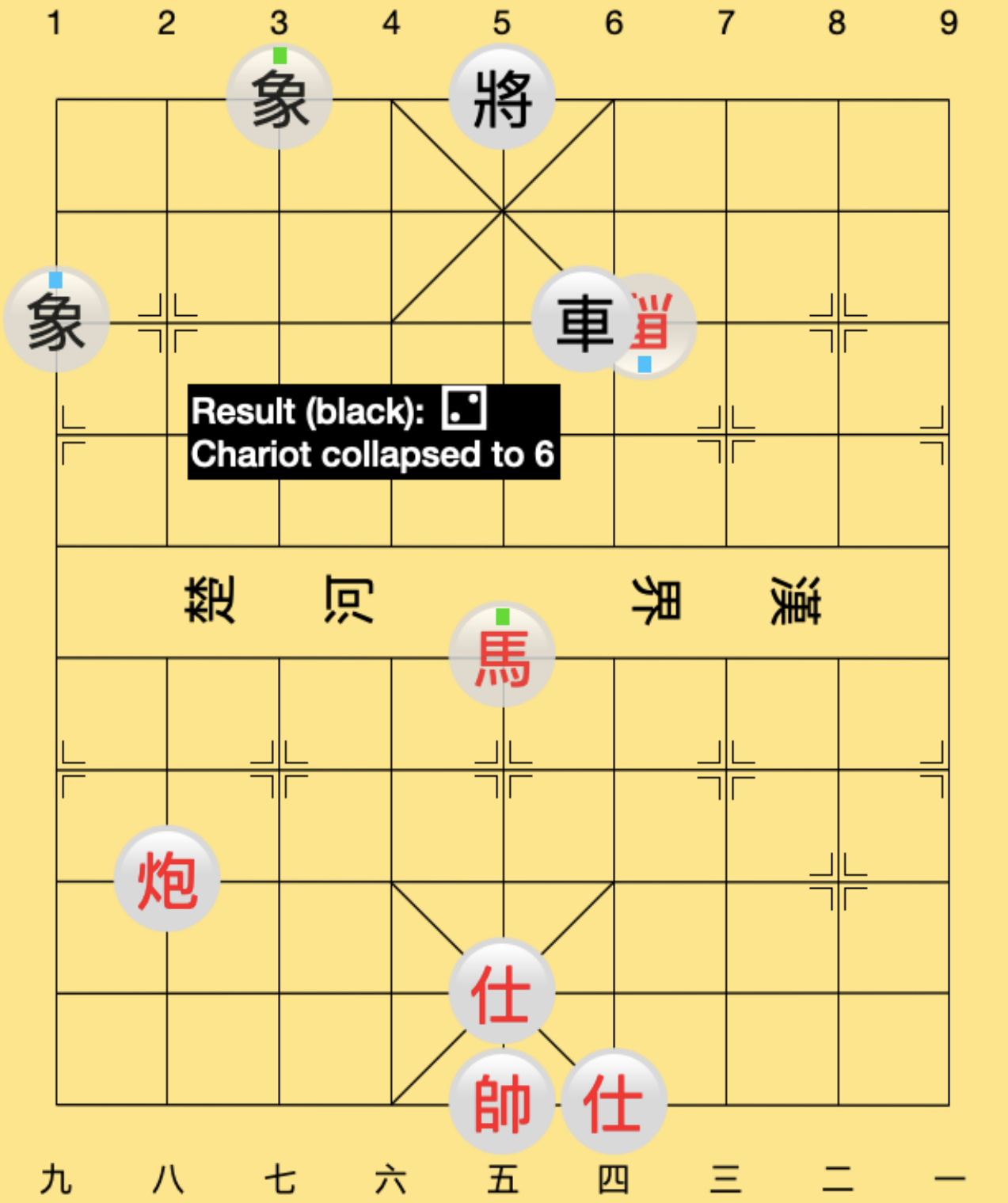}
\vspace*{-0.5cm}
\end{figure}
\begin{figure}[H]
\centering
\includegraphics[width=0.45\columnwidth]{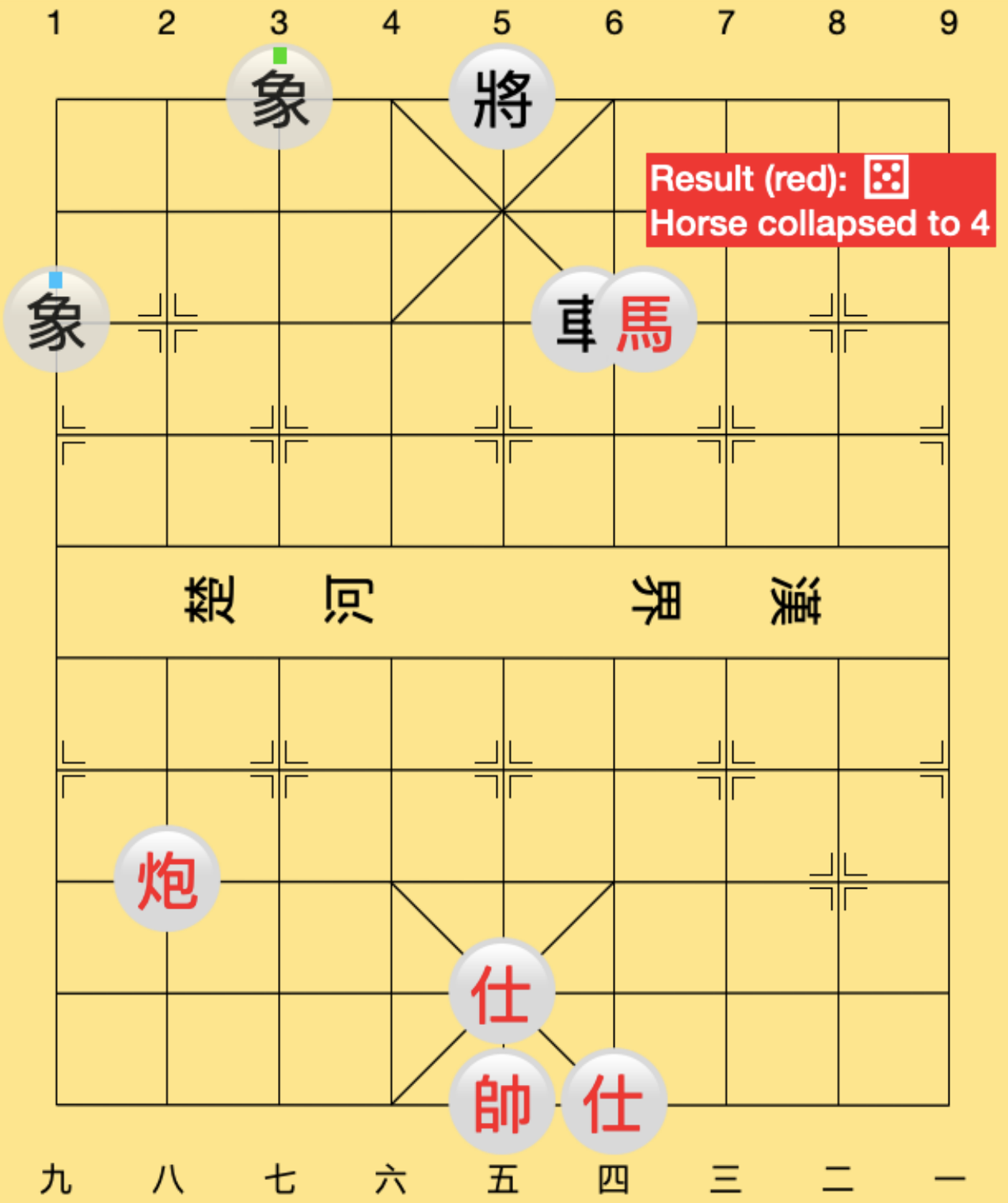}
\end{figure}
\item[3)] If, after step 2, two conventional pieces are on the same intersection, the capture is said to be 'successful': the piece owned by the player attempting the capture stays, while the other is \textbf{captured and removed from the board}.\\
\end{enumerate}
\begin{figure}[H]
\centering
\includegraphics[width=0.45\columnwidth]{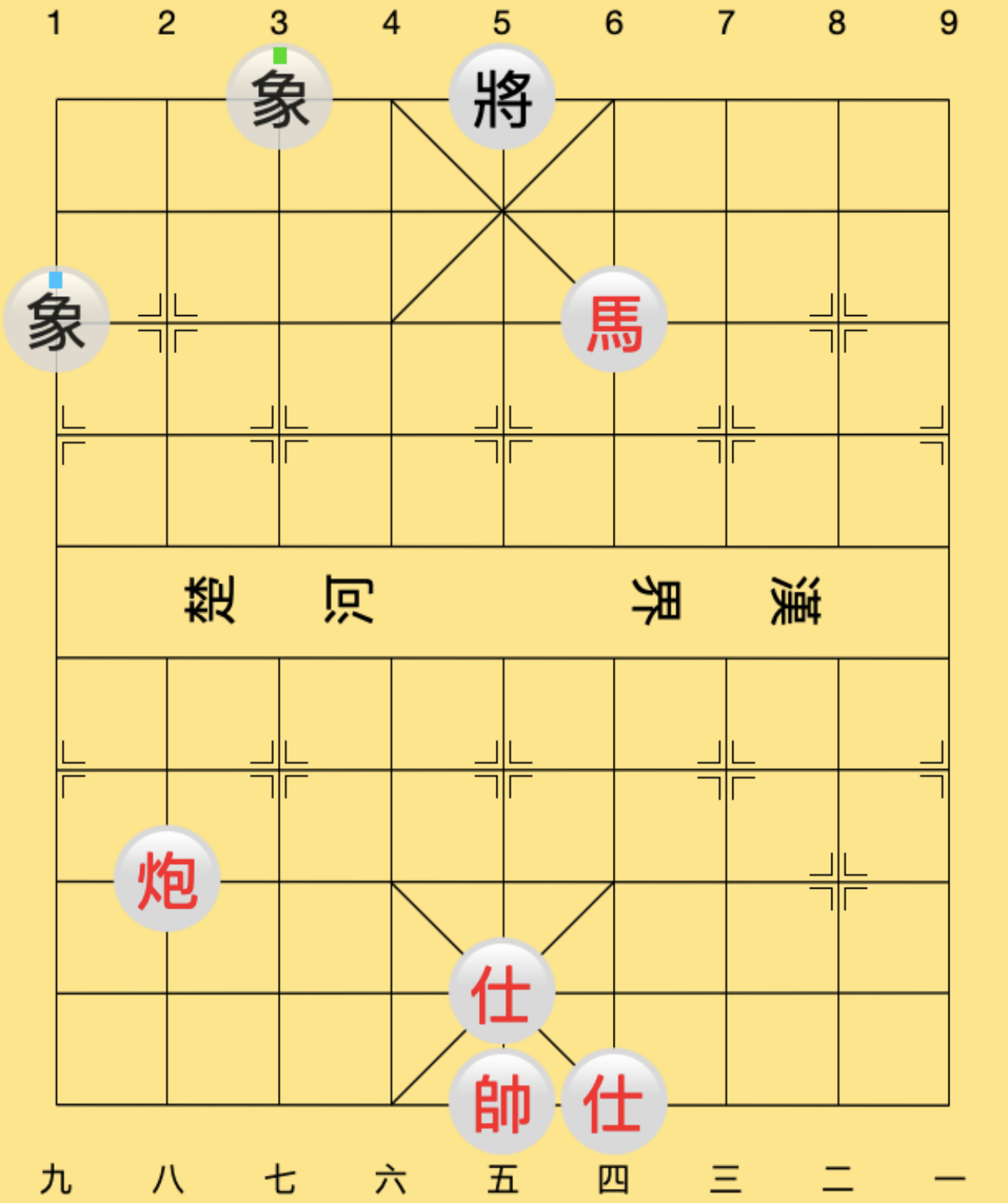}
\end{figure}
\begin{enumerate}[1.1.1]
\item[6.5.1] \textit{[Explanation]} Even the slightest interaction with the environment can cause a superposition state to collapse. That's why, an indefinite piece is meant to be surrounded by a protection layer (sealed box). However, when it hits, or is hit by, another piece, indefinite or not, the isolation gets broken and the superposition state collapses immediately.\\
\end{enumerate}
\item[6.6] A move that could \textbf{potentially} expose the player's King, indefinite or not, to check (by one or more of the opponent's conventional or indefinite pieces), or make that King face the opponent's King, indefinite or not, without any intervening piece, barring Article 7.9, or leave that King in check, counts as 'committing suicide'. (See also Section 5 in \cite{varga}.)
\begin{figure}[H]
\centering
\includegraphics[width=0.45\columnwidth]{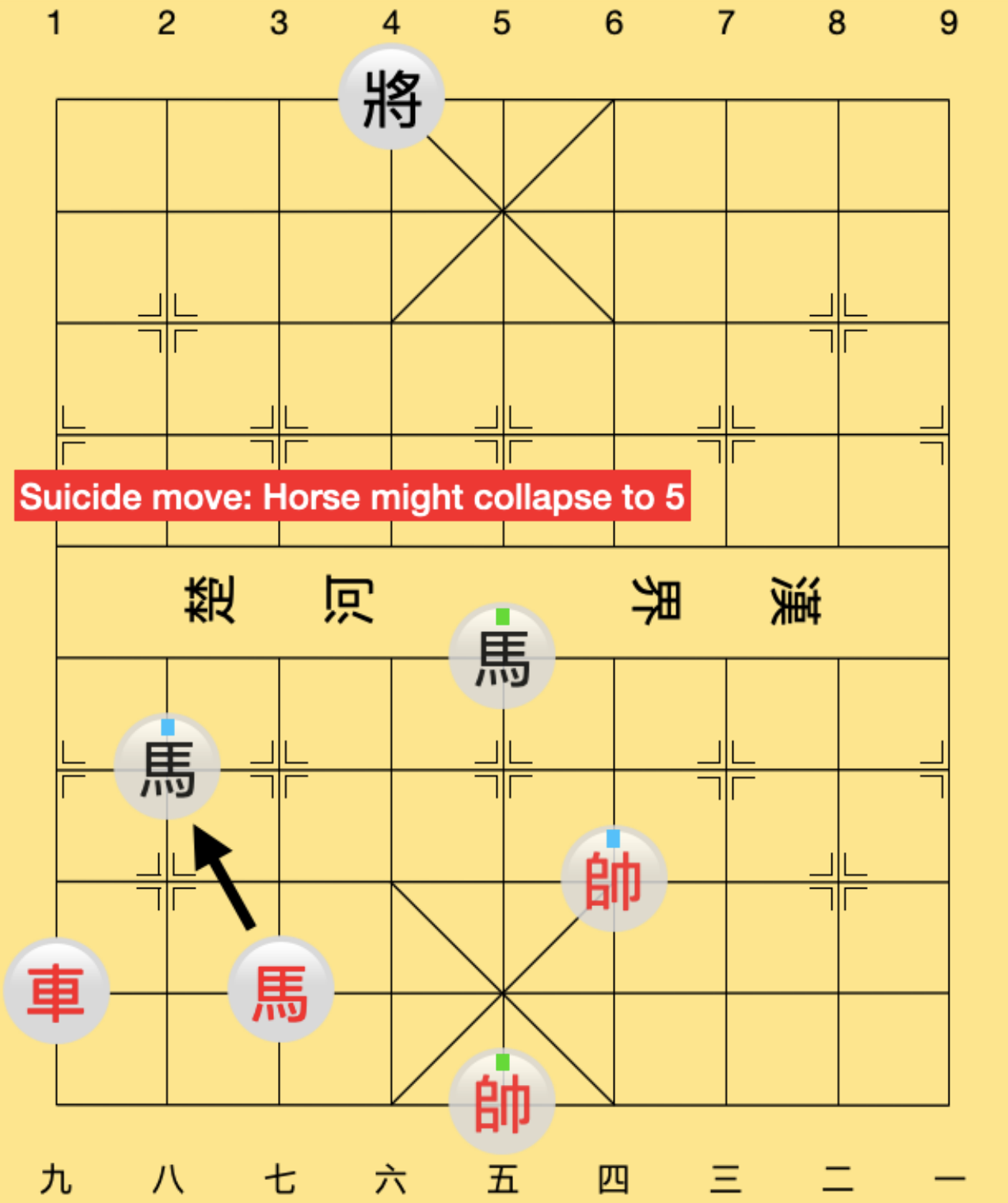}
\end{figure}
\item[6.7] If an indefinite piece is 'one move away' from its paired piece (i.e. it could move there if the intersection on which its paired piece stands was unoccupied, not considering whether such a move would count as committing suicide), then the player who owns the pair may make a move that merely inverts the facing of the marks on both pieces (by rotating each piece by $180^\circ$), provided that it makes a real difference to the position.
\begin{figure}[H]
\centering
\includegraphics[width=0.45\columnwidth]{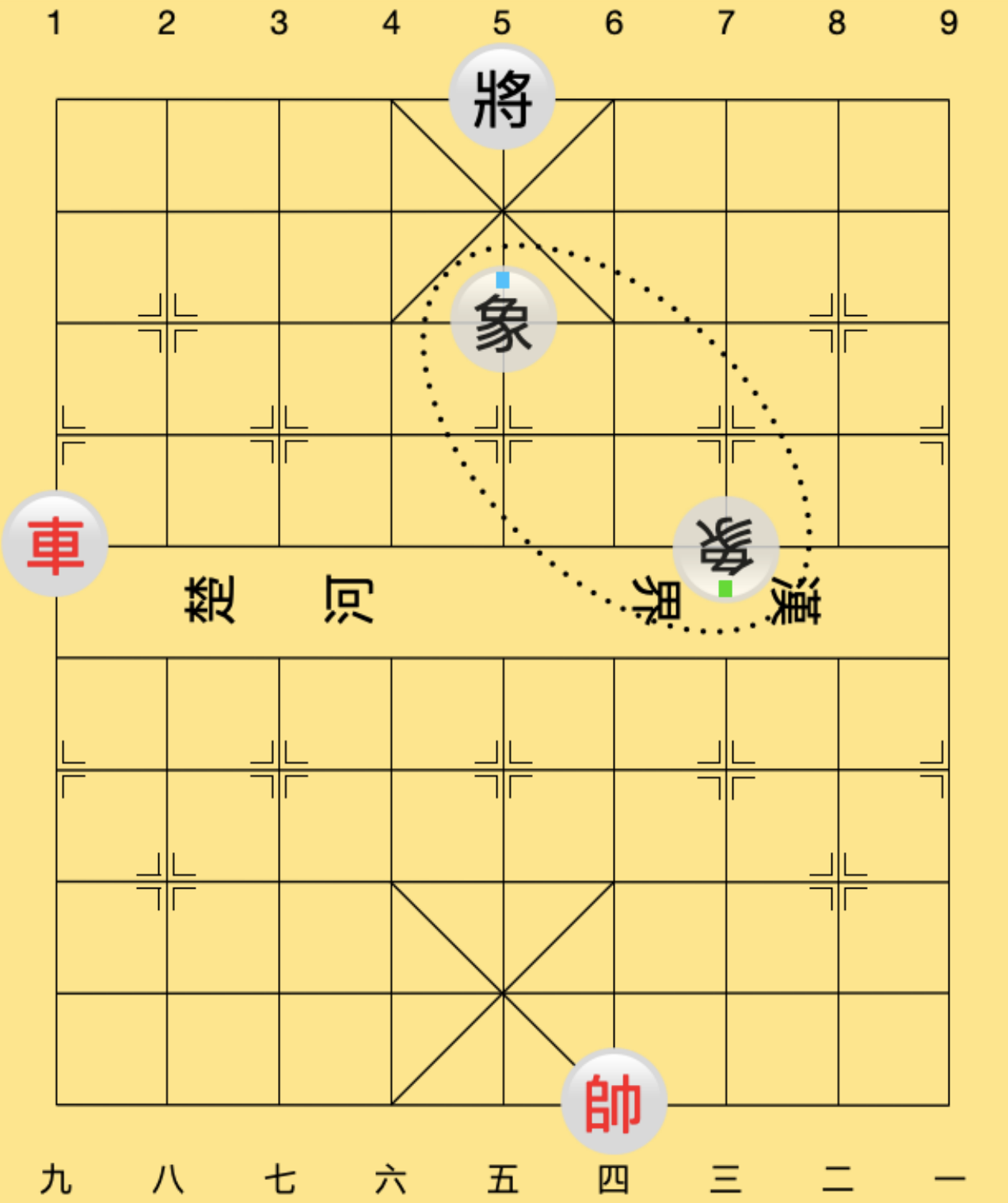}
\vspace*{-0.5cm}
\end{figure}
\begin{figure}[H]
\centering
\includegraphics[width=0.45\columnwidth]{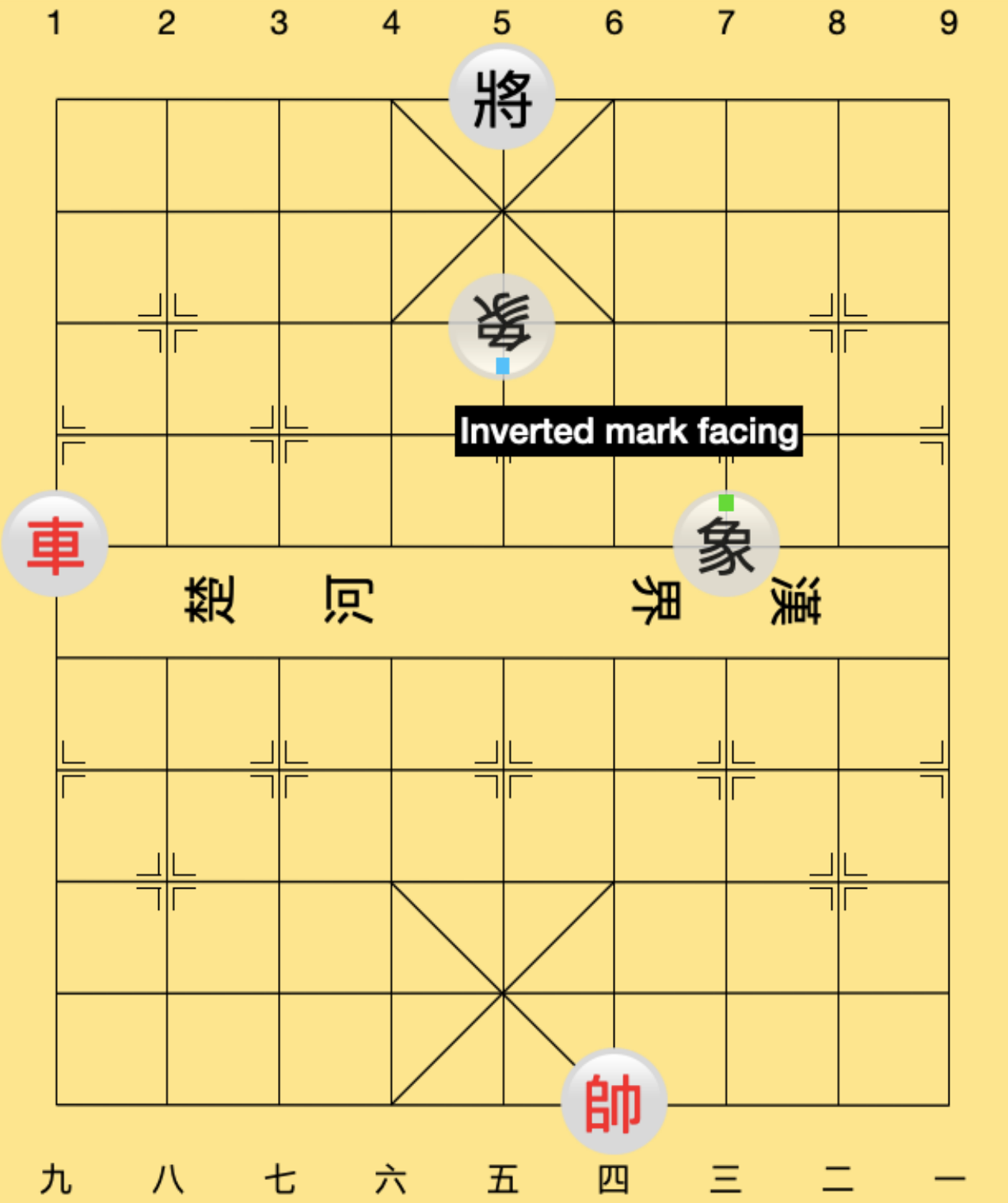}
\end{figure}
\end{enumerate}

\subsection*{Article 7: Entanglement rules}

In case of conflict, Article 7 has priority over Article 6.

\begin{enumerate}[1.1]
\item[7.1] A conventional piece may join the superposition in which an opponent's indefinite piece participates. This is called an 'entanglement move'. The conventional piece has to be replaced by the two pieces of a corresponding indefinite pair instance, one placed on the intersection of departure, and the other on an unoccupied intersection of arrival (reachable as per Articles 2.1 to 2.9). At least one indefinite piece of the pair instance must threaten the opponent's indefinite piece (incl. 'flying capture', as per Article 7.9).\footnote{In the figure below, the dotted ellipse indicates that the Horse on 6 threatens the Chariot on 5. Also, an indefinite Horse on the intersection of arrival 4 would threaten the Chariot on 4. Either of these two reasons is sufficient to allow an entanglement move involving the Chariot pair, as stated in Article 7.1.} (The point of entanglement is that the involved pairs become connected, in the sense that an attempted capture will collapse them in tandem, in a correlated fashion, see Article 7.5.)
\begin{figure}[H]
\centering
\includegraphics[width=0.45\columnwidth]{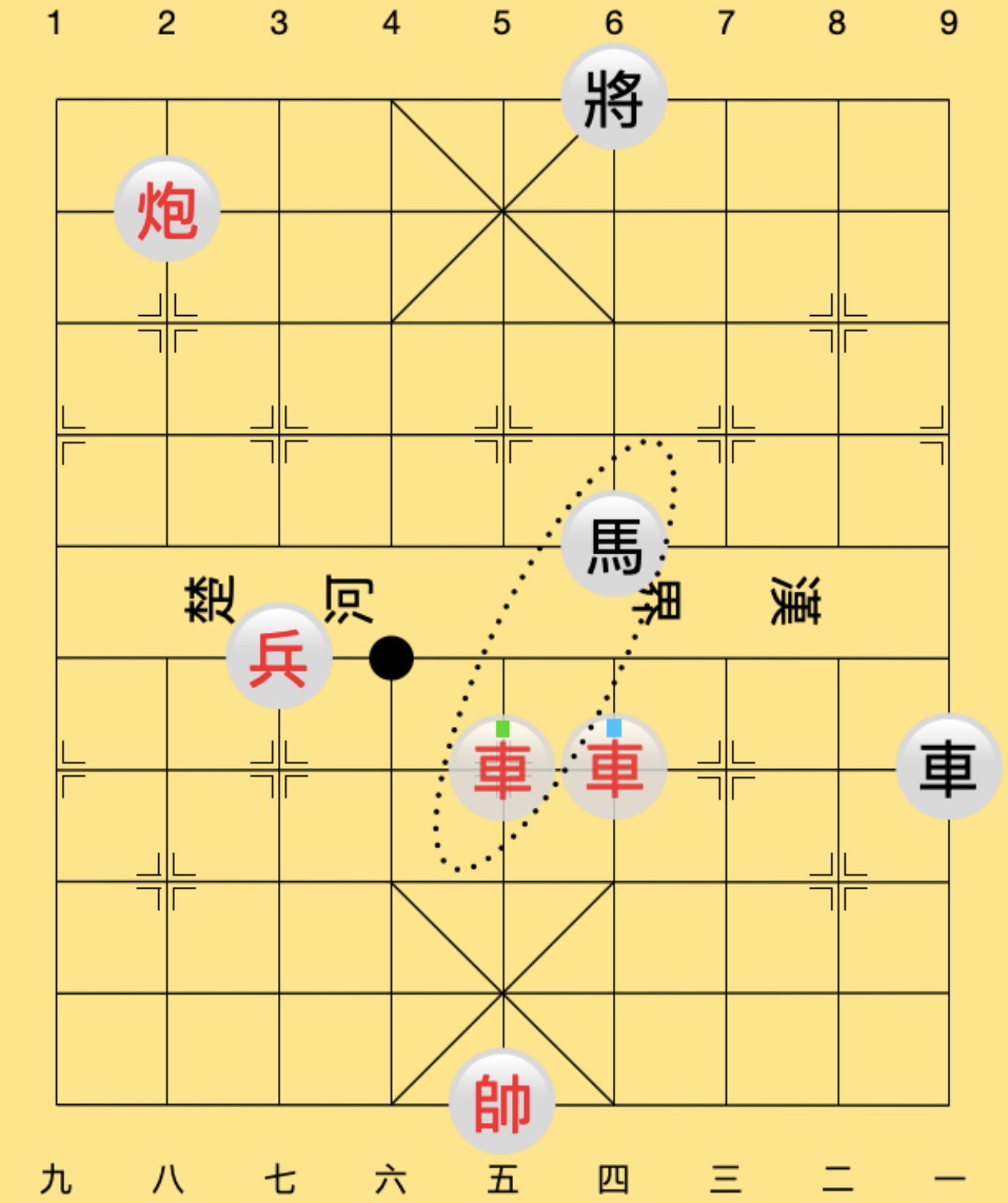}
\end{figure}
\item[7.2] For Horse, Cannon and Chariot, an entanglement move is allowed only if the intersection of arrival is on the opposite side of the River from the departure. For Adviser, Elephant and King, the precondition for an entanglement move is that the intersection of departure belongs to the Palace. For Pawn, it may make an entanglement move only after it has crossed the River.\\
\item[7.3] The mark on each indefinite piece which hasn't yet been involved in an entanglement move must lie on a file (just like in all figures so far).\\
\item[7.4] Steps to direct the marks in an entanglement move:\\
\begin{enumerate}[1.]
\item[1)] The player making the move directs the marks of his/her newly placed indefinite pair instance such that they both lie on a file and indicate \textbf{the same type of superposition} (equal or unequal) as those of the opponent's pair involved.
\begin{figure}[H]
\centering
\includegraphics[width=0.45\columnwidth]{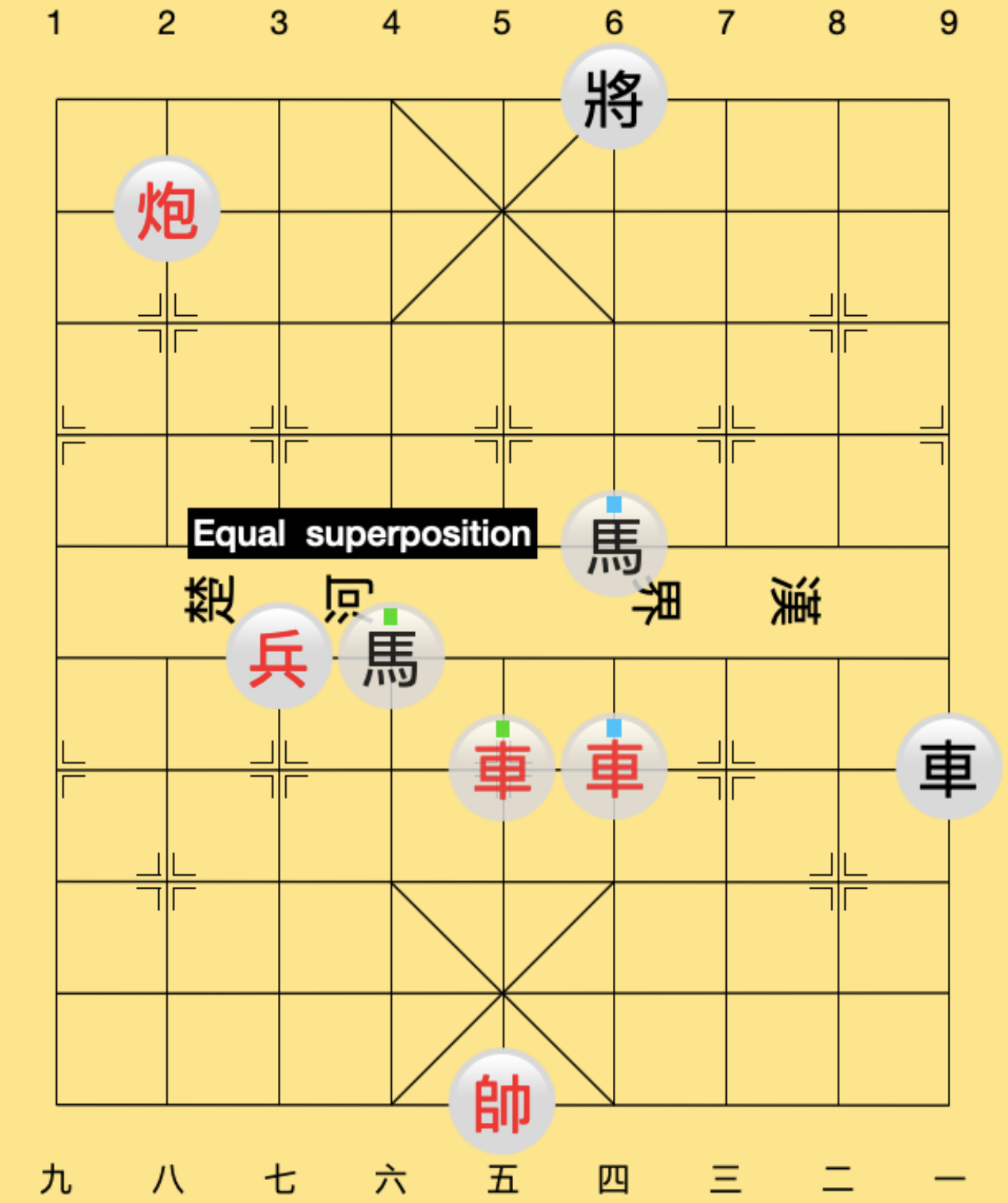}
\end{figure}
\item[2)] If the opponent's marks lie on files, the opponent must \textbf{rotate both pieces by the same angle}, either $+45^\circ$ or $-45^\circ$, to make the marks lie on the same diagonal (up to a parallel shift). A diagonal cannot be chosen if there are already other indefinite pieces whose marks lie on the same diagonal. If neither of the diagonals can be chosen, the entanglement move \textbf{is not allowed}.
\begin{figure}[H]
\centering
\includegraphics[width=0.45\columnwidth]{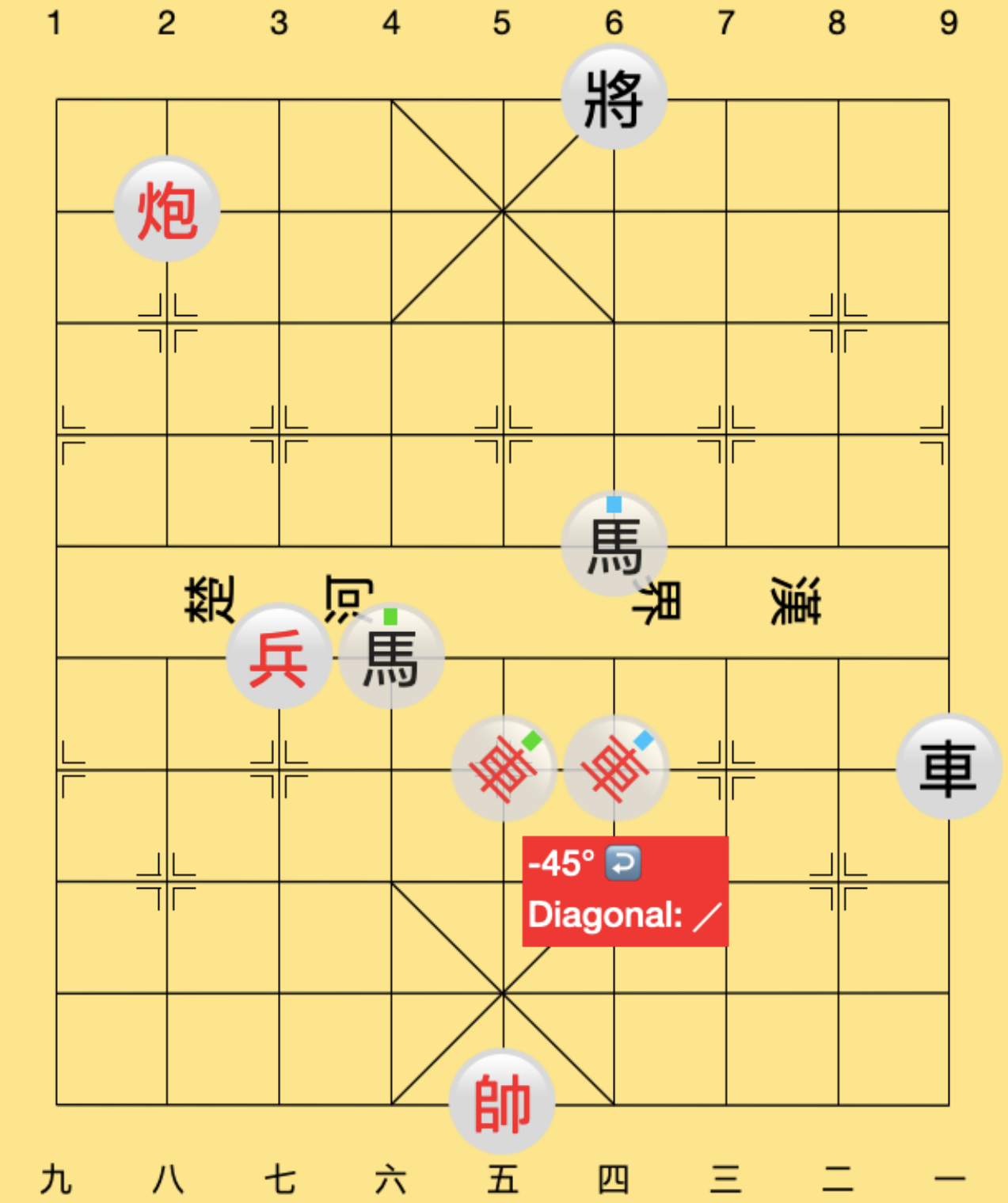}
\end{figure}
\item[3)] The player making the move rotates his/her pieces to align their marks with those of the opponent's.
\begin{figure}[H]
\centering
\includegraphics[width=0.45\columnwidth]{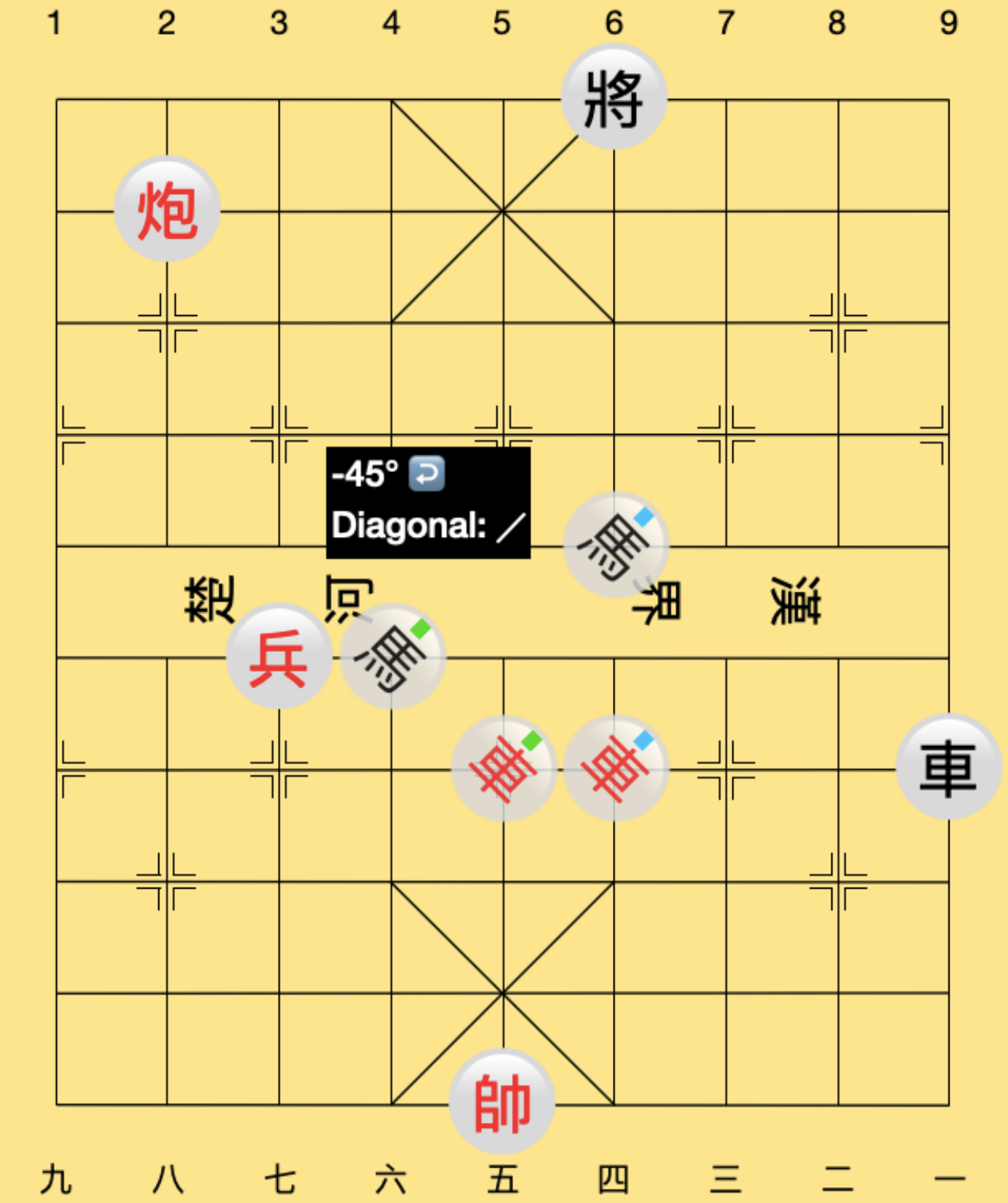}
\end{figure}
\end{enumerate}
\begin{enumerate}[1.1.1]
\item[7.4.1] \textit{[Clarification]} In accordance with Article 6.2, the rotations don't change the type of any superposition, as every mark will face the same upper or lower side of the board as before.\\
\item[7.4.2] \textit{[Explanation]} To justify the limitation imposed by step 2, it is noted that in quantum information, entanglement is a \textbf{resource}, which can be scarce. That's why, players are allowed only limited access to it.\\
\item[7.4.3] \textit{[Clarification]} Additional conventional pieces may join as well (with step 2 omitted).
\begin{figure}[H]
\centering
\includegraphics[width=0.45\columnwidth]{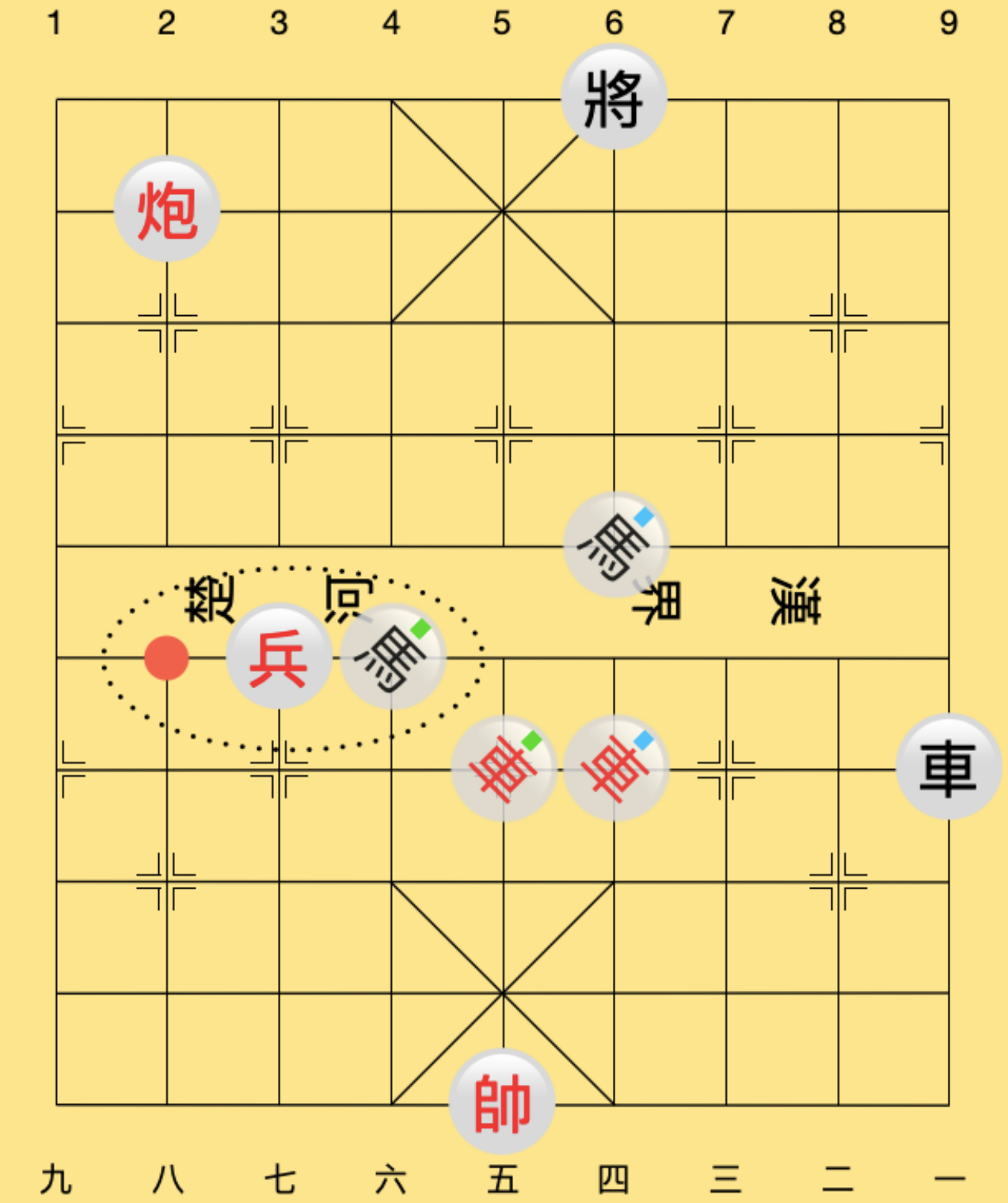}
\vspace*{-0.5cm}
\end{figure}
\begin{figure}[H]
\centering
\includegraphics[width=0.45\columnwidth]{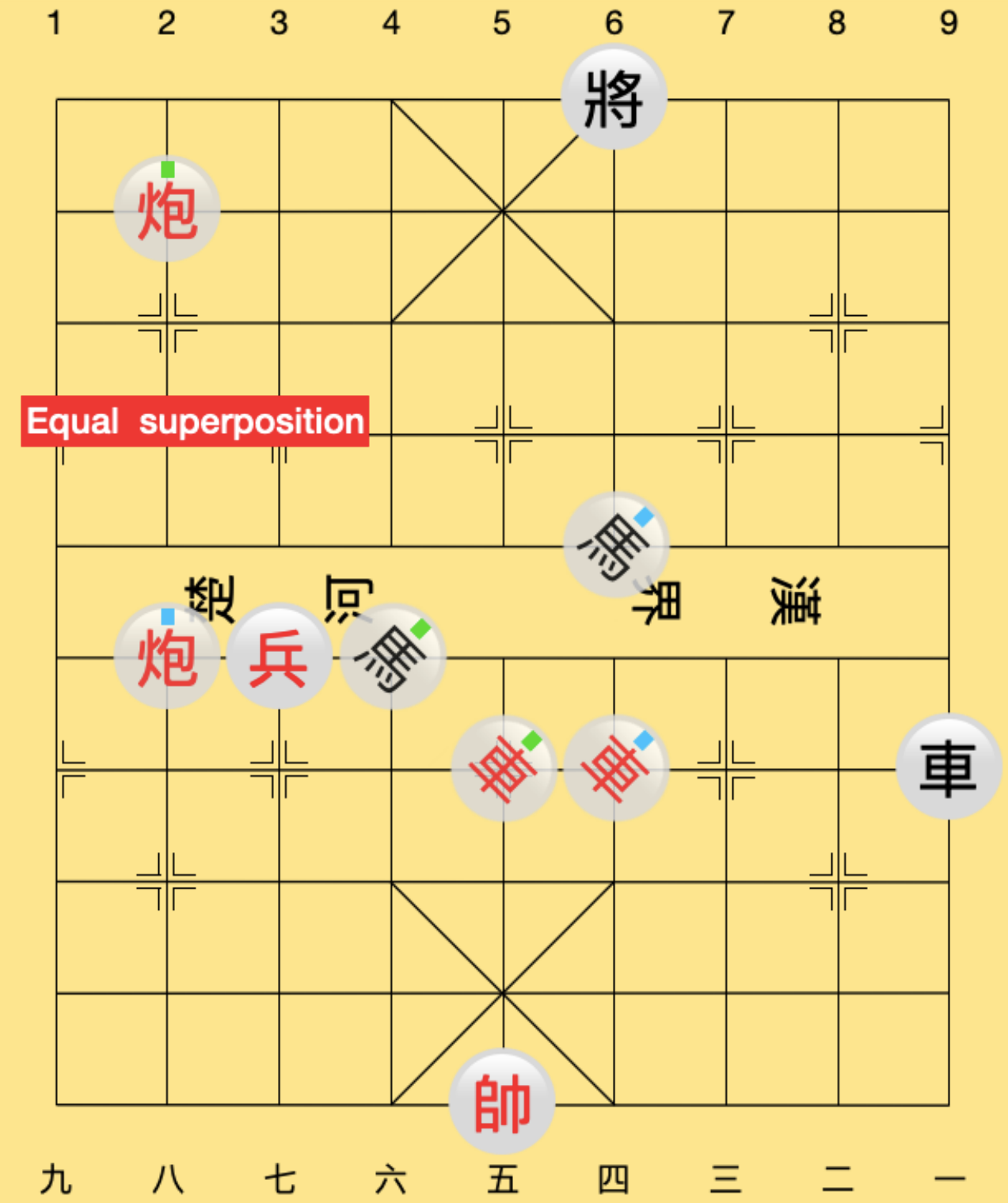}
\vspace*{-0.5cm}
\end{figure}
\begin{figure}[H]
\centering
\includegraphics[width=0.45\columnwidth]{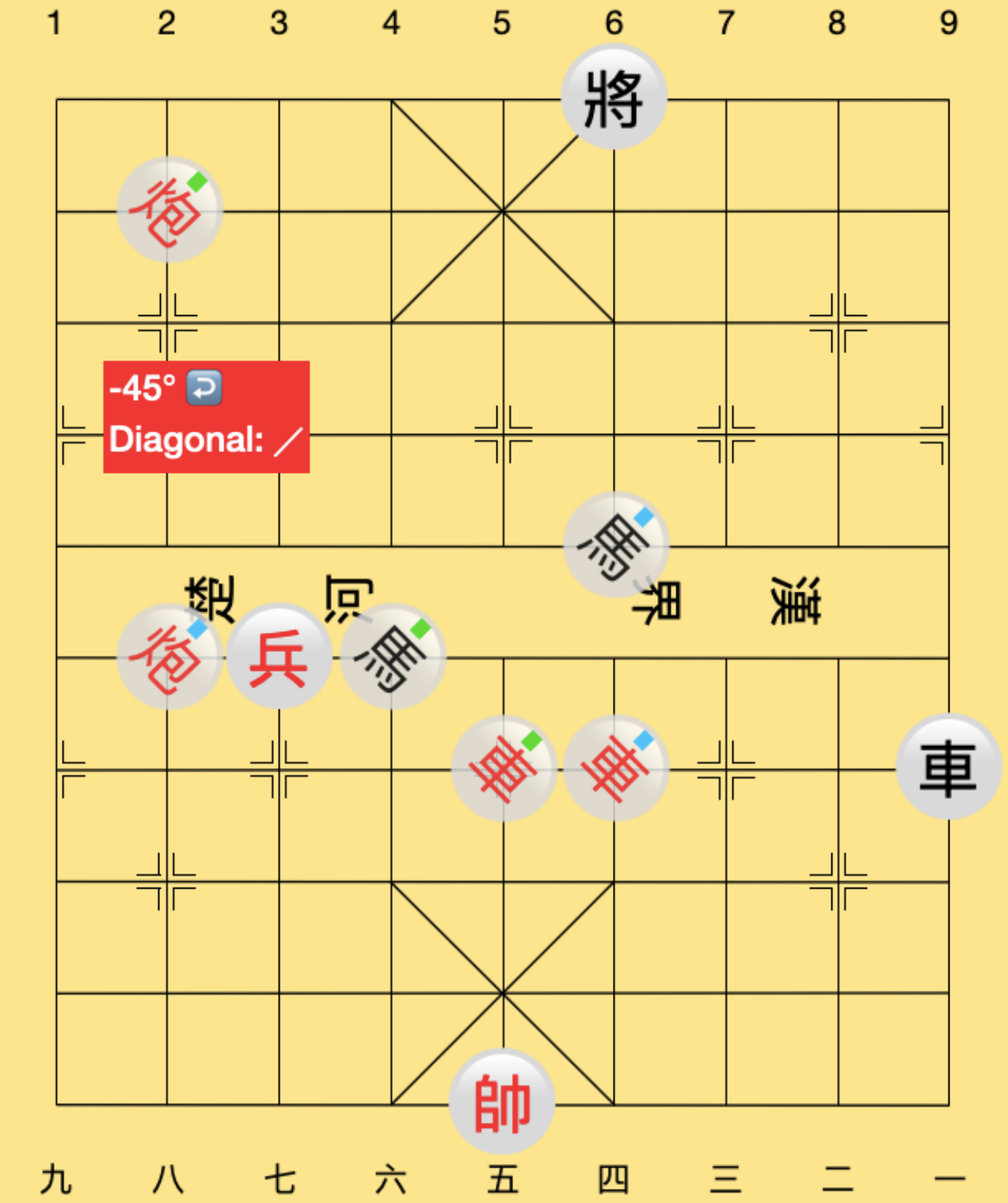}
\end{figure}
\item[7.4.4] \textit{[Clarification]} The indefinite pair instances whose marks lie on the same diagonal are said to be 'entangled' (as they represent the corresponding group of conventional pieces in an entangled state of correlated behaviour, see Article 7.5). Since there are only two diagonals, there can be at most two entangled groups of pairs on the board. (Using a more complex scheme of directing the marks would enable the maximum of $16$ groups of entangled pairs. However, it would likely reduce the playability of the game.)
\begin{figure}[H]
\centering
\includegraphics[width=0.45\columnwidth]{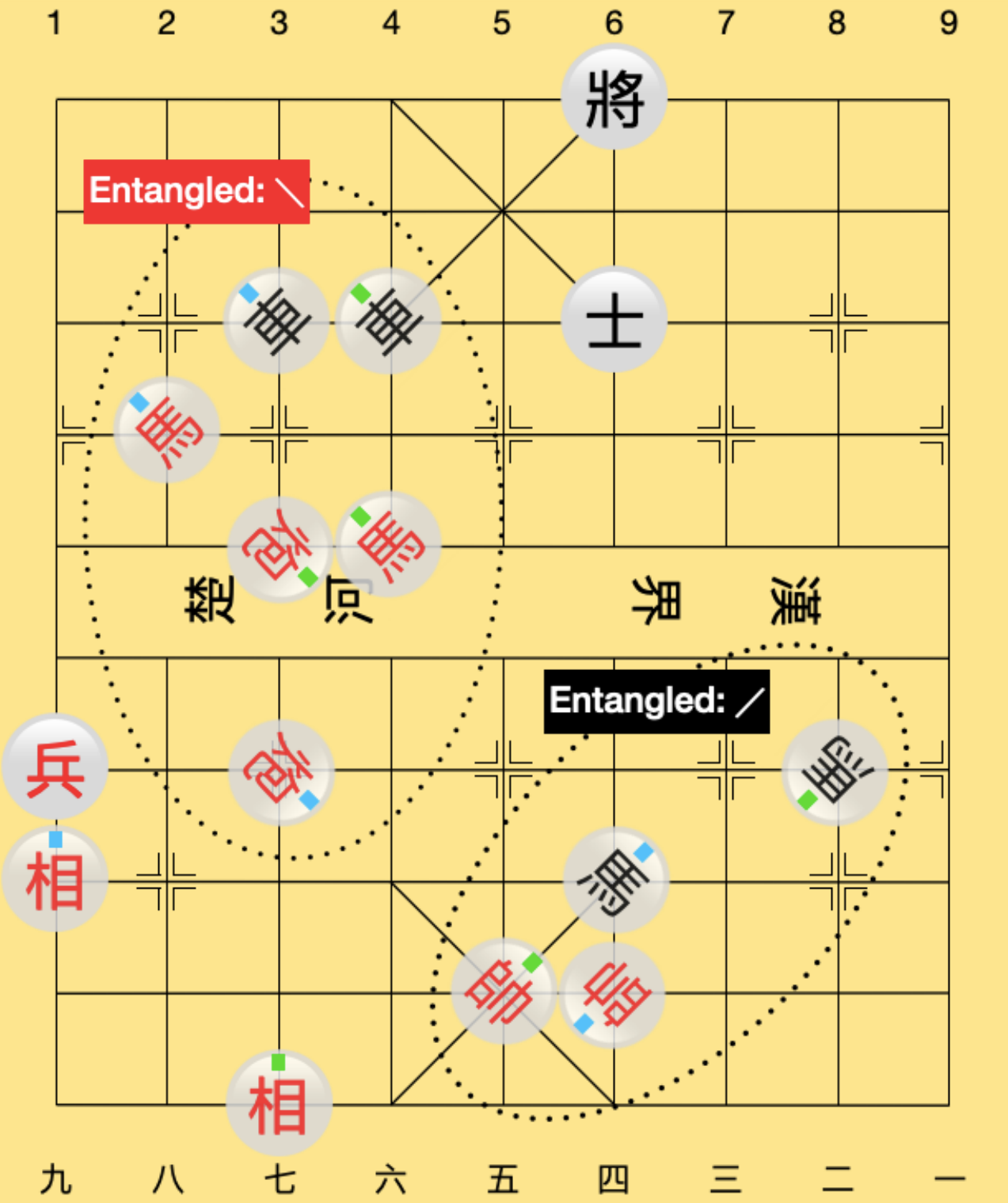}
\end{figure}
\end{enumerate}
\item[7.5] In step 2 of Article 6.5, right after collapsing an indefinite pair following a roll of the dice, all indefinite pair instances entangled with that pair must be collapsed as well, using the same result of the roll.\\
\begin{enumerate}[1.1.1]
\item[7.5.1] \textit{[Explanation]} Thus, rolling the dice results (randomly) in one of two possible collapse outcomes for the whole group of entangled pairs. This is because physically, \textbf{entanglement is a superposition} of those two possible holistic outcomes.
\end{enumerate}
\begin{figure}[H]
\centering
\includegraphics[width=0.45\columnwidth]{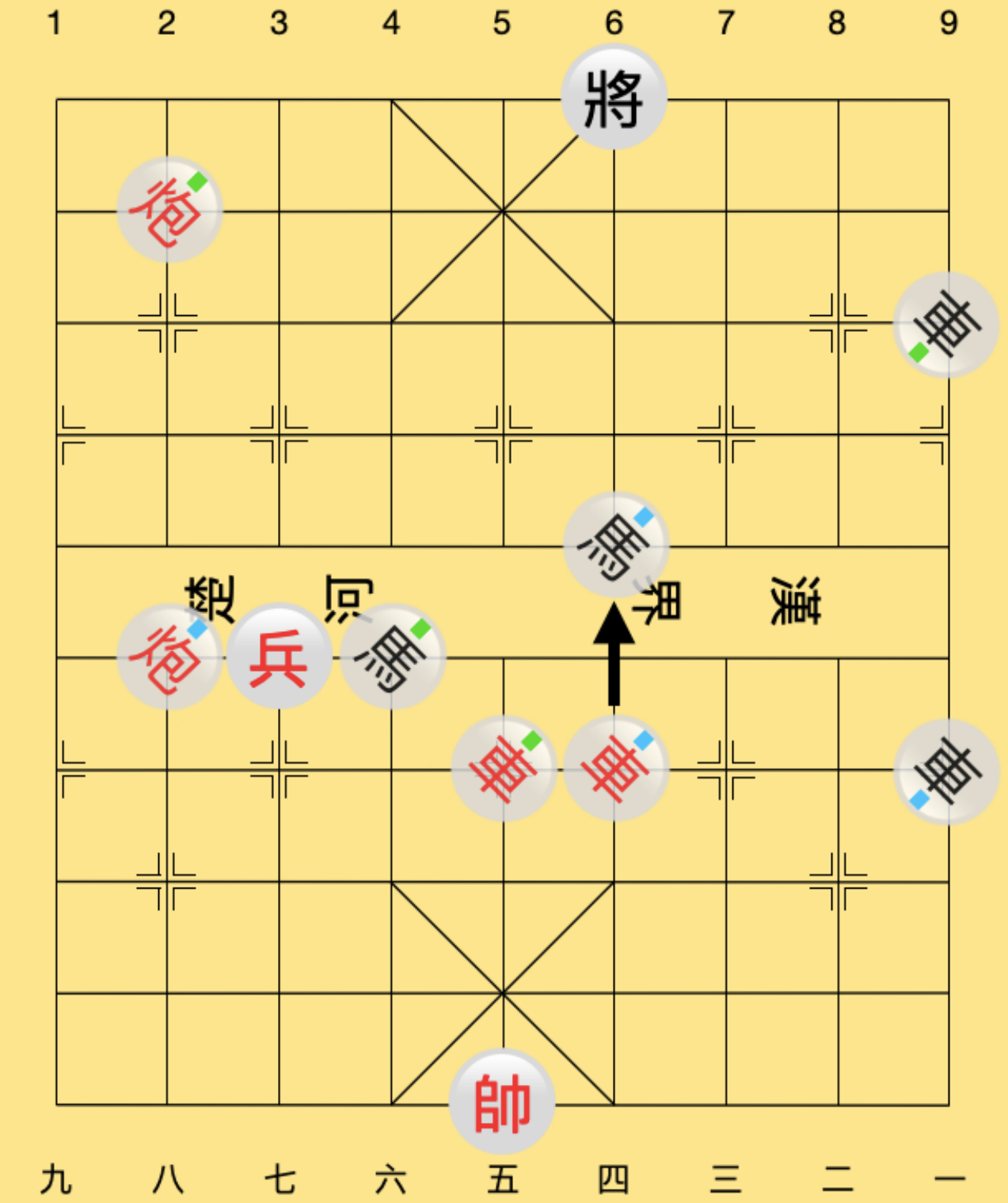}
\vspace*{-0.5cm}
\end{figure}
\begin{figure}[H]
\centering
\includegraphics[width=0.45\columnwidth]{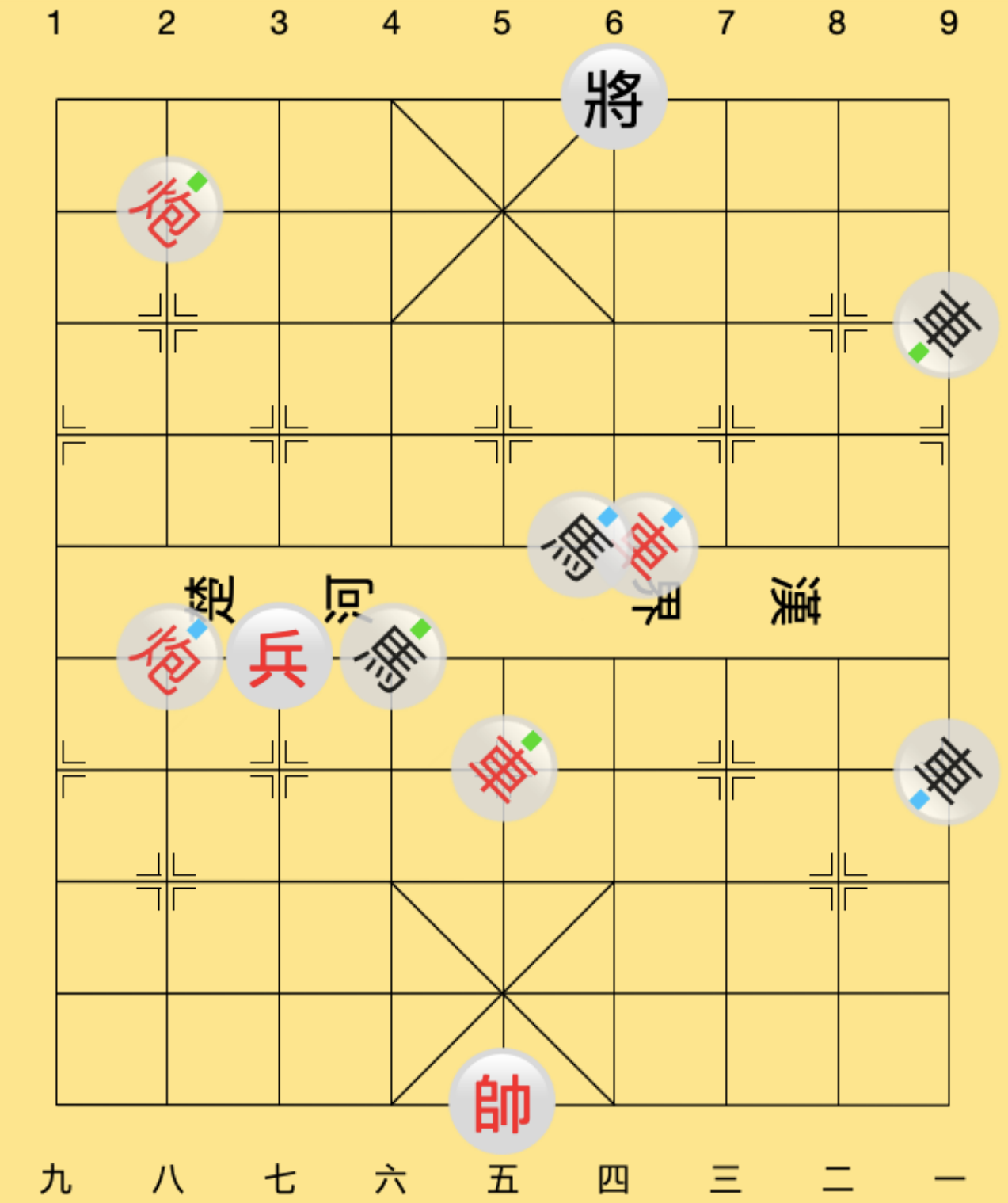}
\vspace*{-0.5cm}
\end{figure}
\begin{figure}[H]
\centering
\includegraphics[width=0.45\columnwidth]{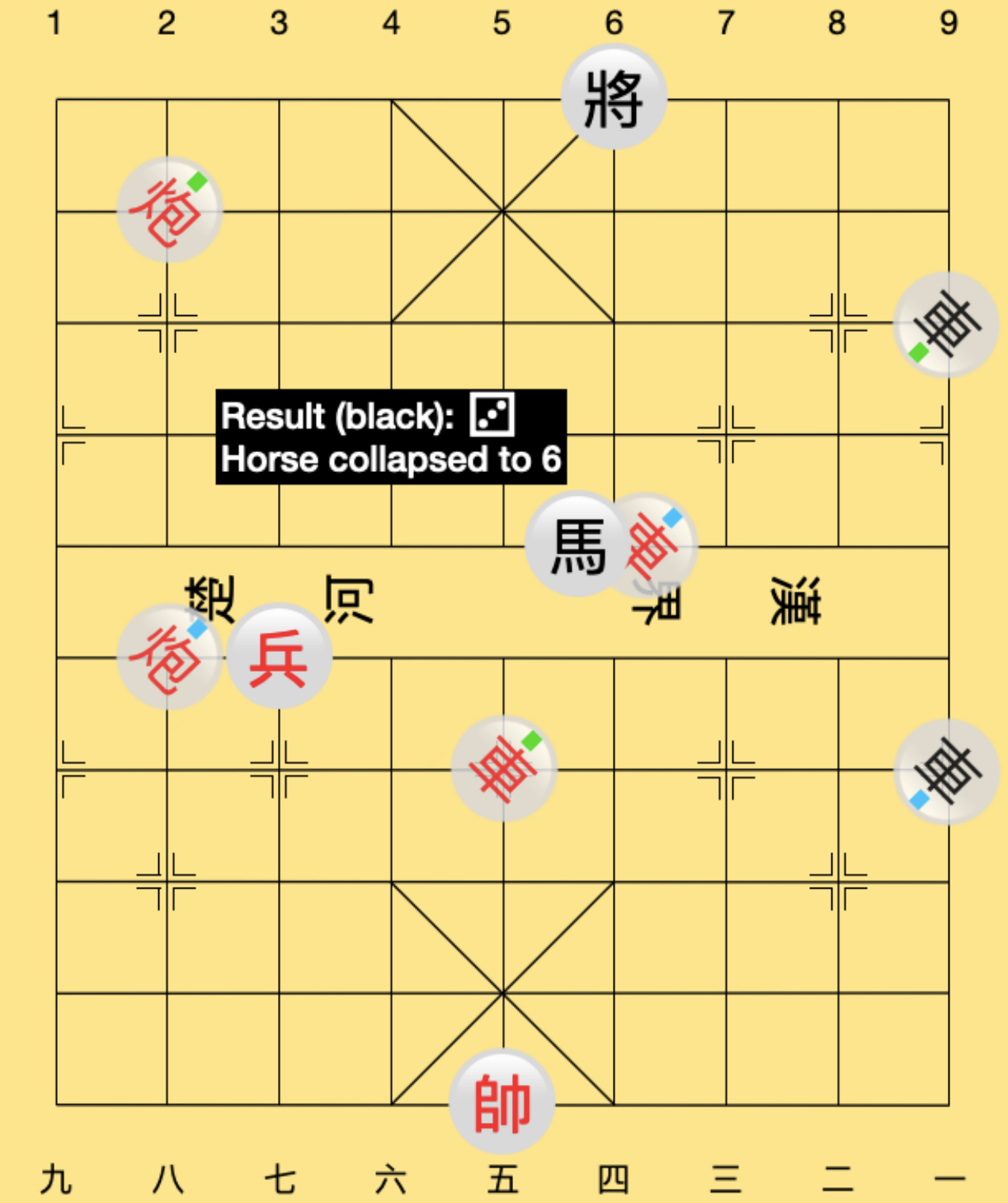}\hfill\includegraphics[width=0.45\columnwidth]{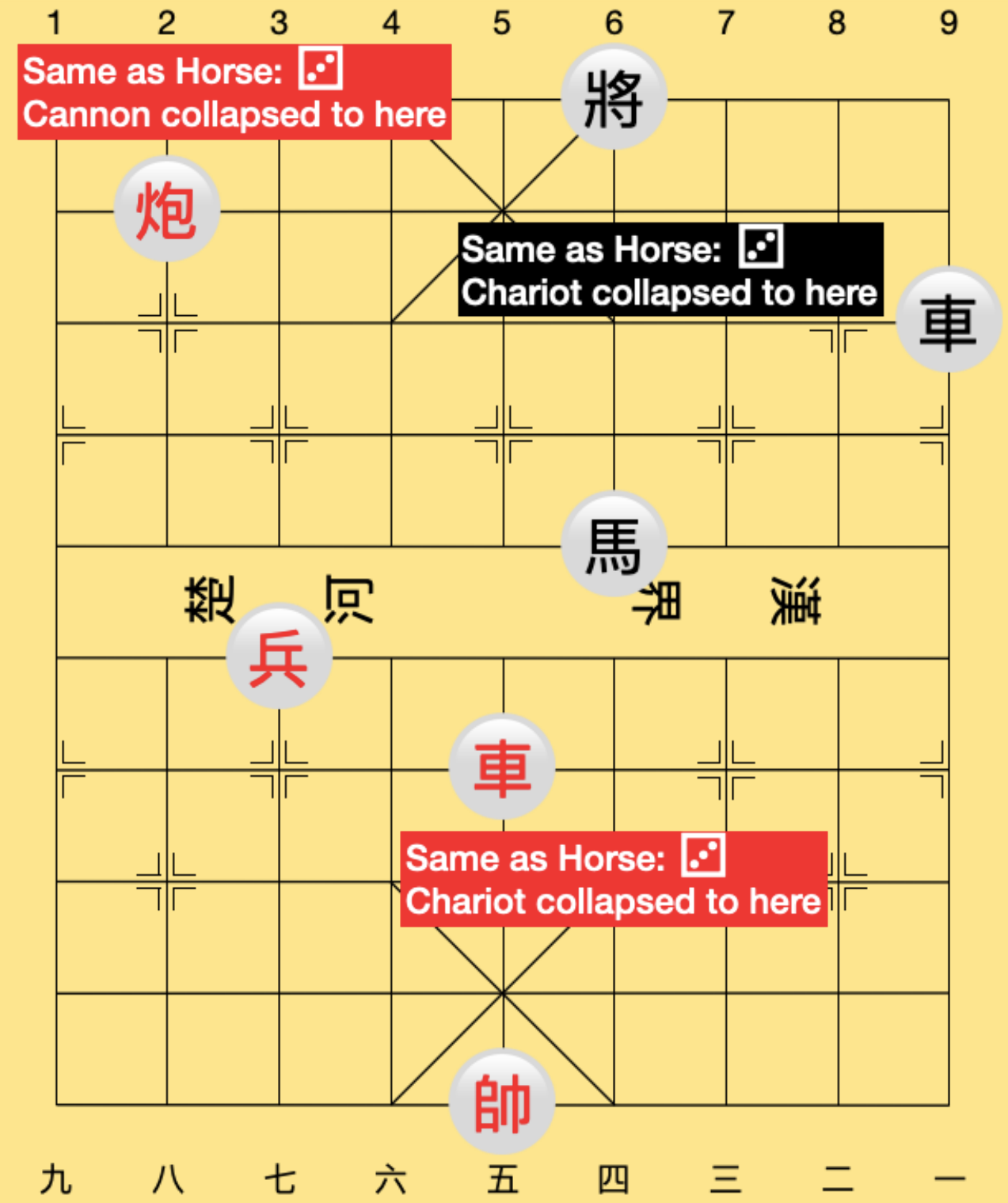}
\end{figure}
\item[7.7] \textit{[Clarification]} The inverting of the facing of the marks as described in Article 6.7 (i.e. rotating the pieces by $180^\circ$) can be performed also with an indefinite pair instance that is entangled with one or more other indefinite pair instances.
\begin{figure}[H]
\centering
\includegraphics[width=0.45\columnwidth]{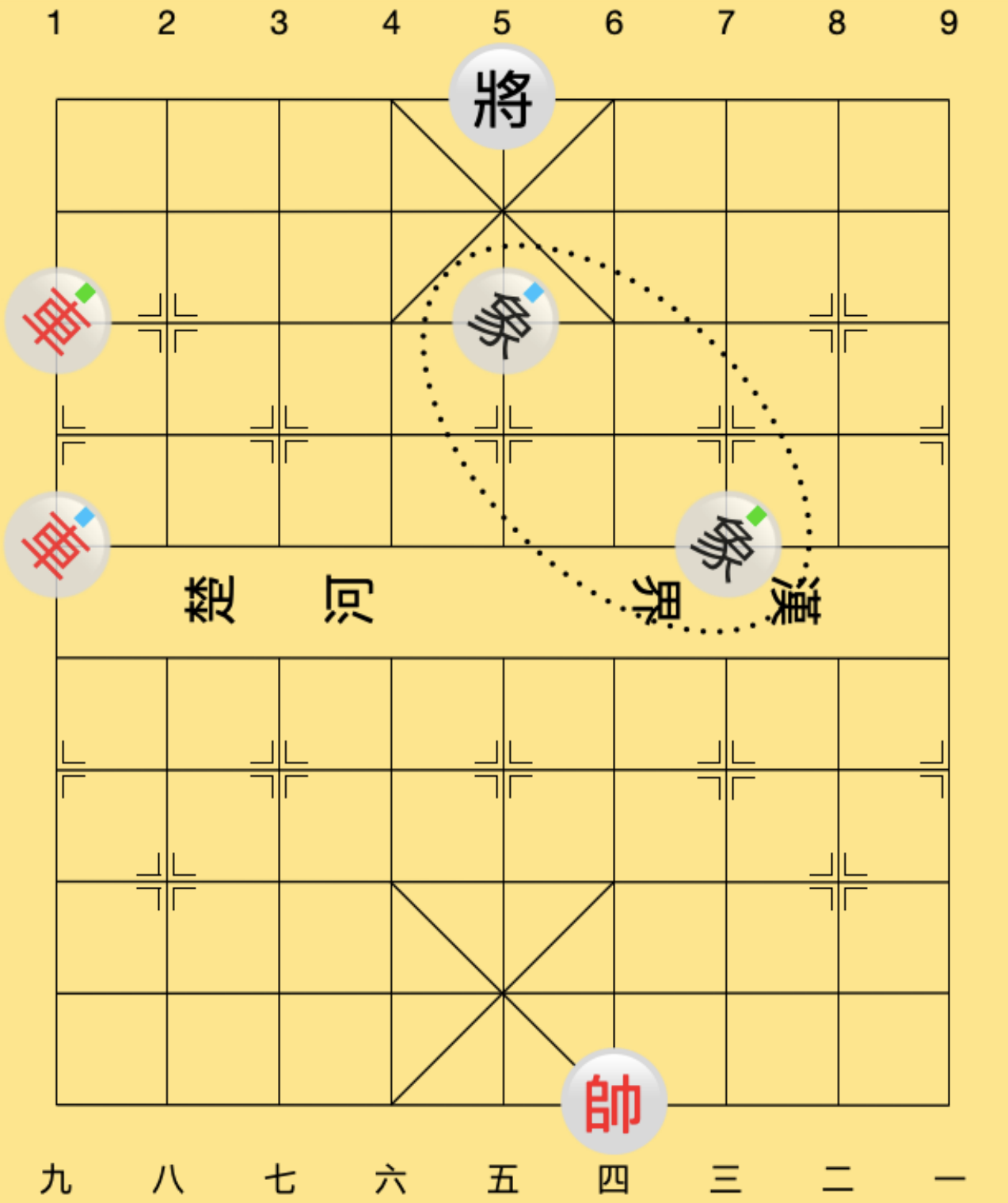}
\vspace*{-0.5cm}
\end{figure}
\begin{figure}[H]
\centering
\includegraphics[width=0.45\columnwidth]{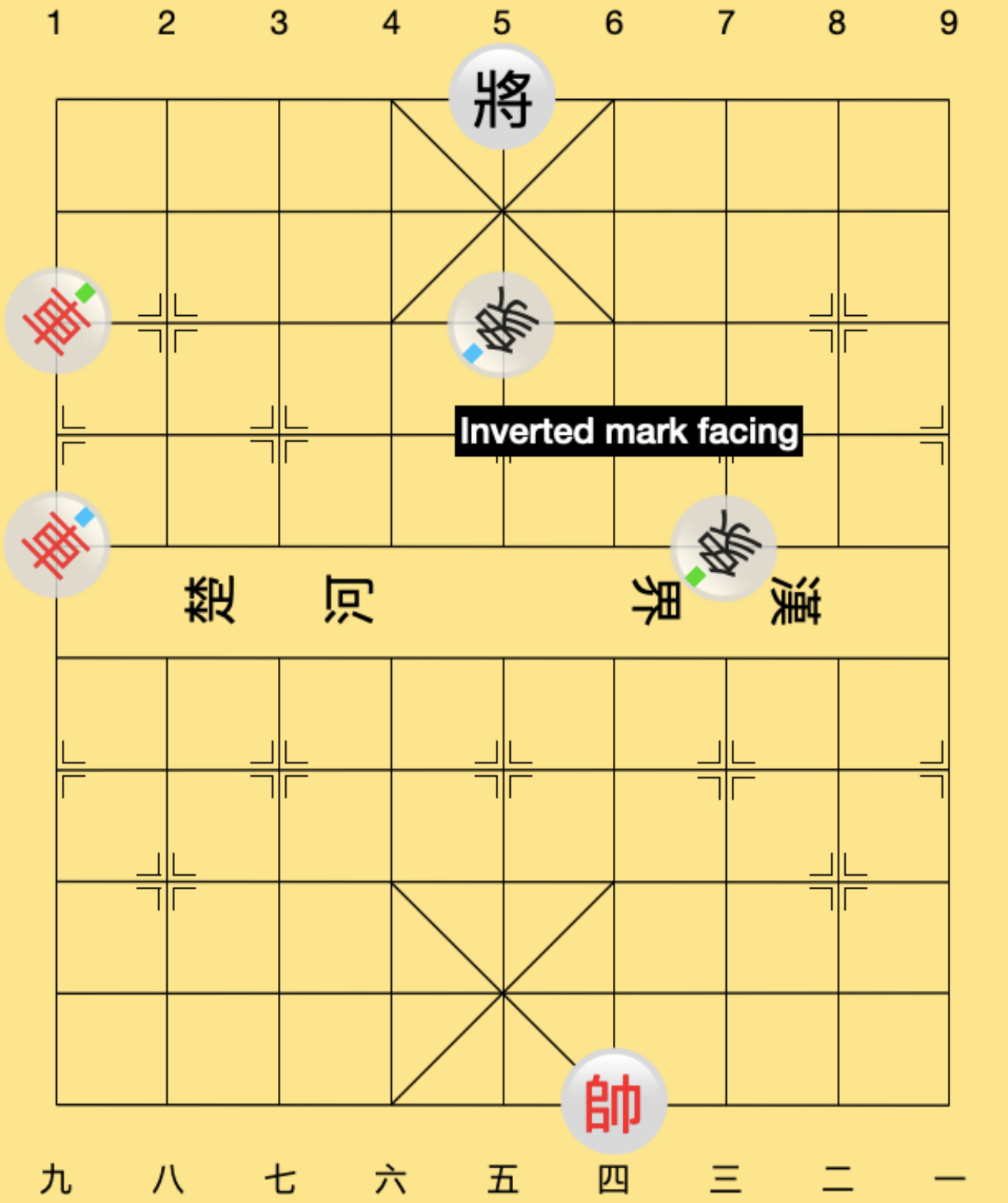}
\end{figure}
\item[7.8] If the King, indefinite or not, is threatened by an opponent's piece, it counts as a 'check' only if a hypothetical capture of that King piece would have a greater than zero chance of success.\\
\item[7.9] The King, indefinite or not, is allowed to be left under, or to be exposed to, threat by an opponent's piece (incl. 'flying capture'), provided there is zero chance of a successful capture of it. That is, it does not count as committing suicide.\\
\item[7.10] Attempting to capture the opponent’s King (excl. 'flying capture'), indefinite or not, is allowed if it has zero chance of success.
\begin{figure}[H]
\centering
\includegraphics[width=0.45\columnwidth]{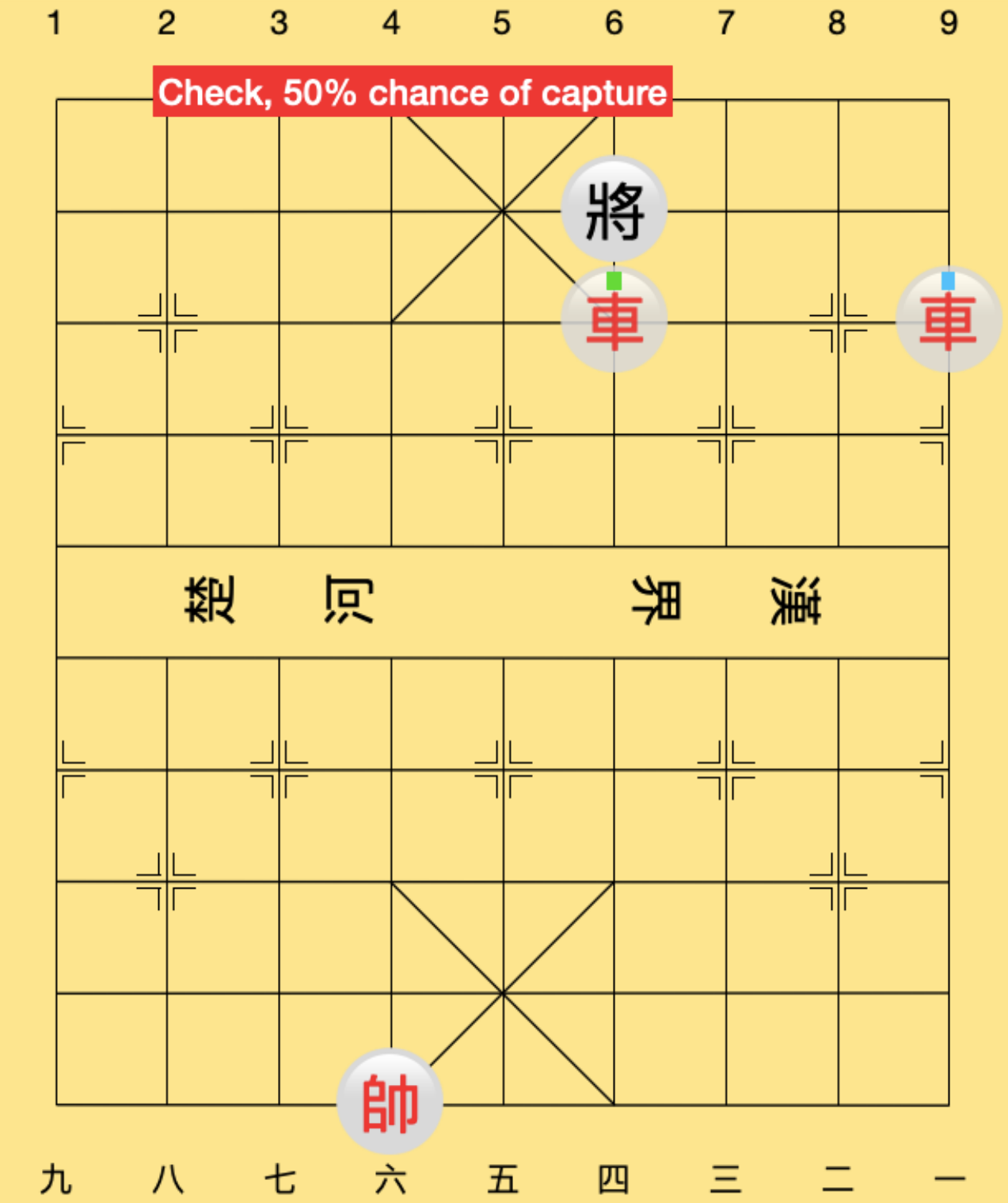}
\vspace*{-0.5cm}
\end{figure}
\begin{figure}[H]
\centering
\includegraphics[width=0.45\columnwidth]{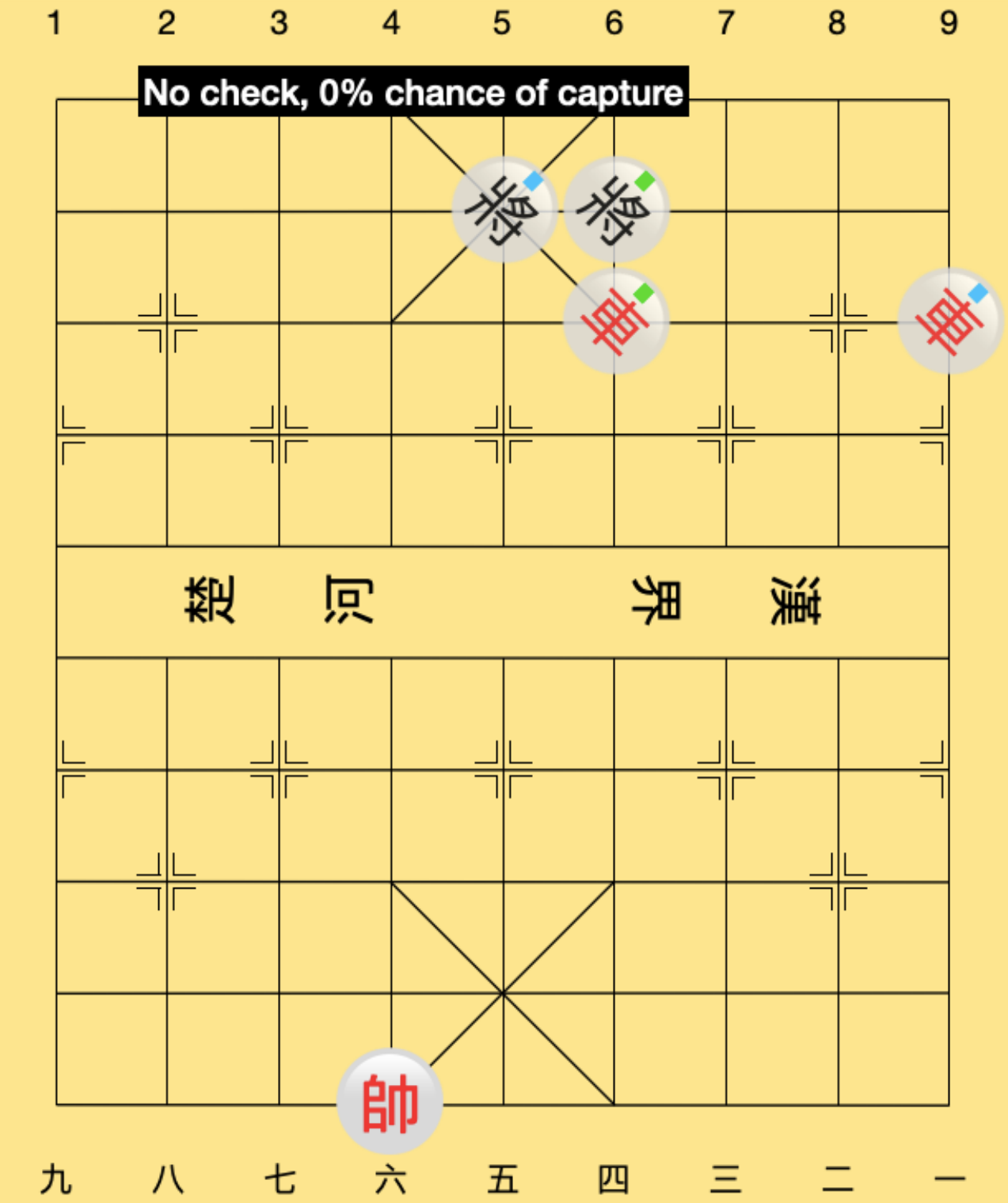}
\end{figure}
\end{enumerate}

\section{Competitive rules of play}\label{competitive}

The detailed specification of competitive rules is out of scope in this paper.

A couple of possible adjustments to the basic rules, such as using a quantum random number generator (QRNG) \cite{quantis} instead of rolling the dice, slowing down the superposition move, or completely eliminating randomness from the game, are listed in \cite{varga}.

\section*{Appendix A: Other variants}\label{variants}

Plenty of potential variants are mentioned in the Appendices of \cite{varga} which would also work with Xiangqi. A Xiangqi-specific variant could be if the indefinite Cannon wouldn't be allowed to use its pair piece as a Cannon mount. Alternatively, the game could be made more complex by dropping Articles 6.3 and 7.2.

The flexible system of marks and rotations enables players and educators alike to come up with new rules to illustrate additional quantum phenomena, such as quantum interference as described in Appendix D in \cite{varga}, or partial collapse by allowing conventional pieces to be on three or more intersections at once, or complex amplitudes via fine-grained rotations, or even quantum tunnelling. The limit is the imagination.

An example of tunnelling is shown below: by rolling a 6, the Horse successfully tunnelled through the Cannon to an unoccupied intersection. Otherwise, the Horse would have "bounced back" and stayed where it was.

\begin{figure}[H]
\centering
\includegraphics[width=0.45\columnwidth]{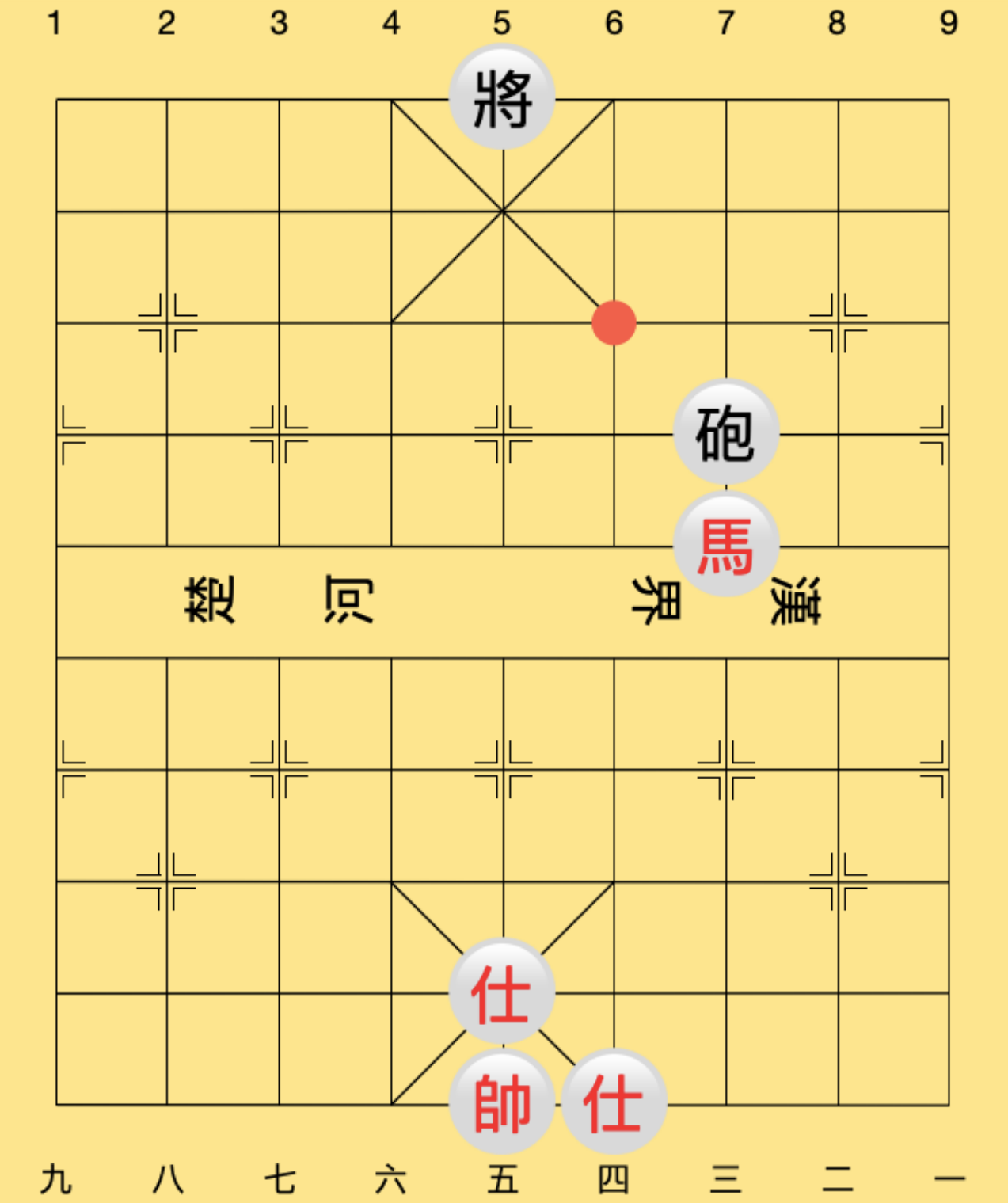}
\vspace*{-0.5cm}
\end{figure}
\begin{figure}[H]
\centering
\includegraphics[width=0.45\columnwidth]{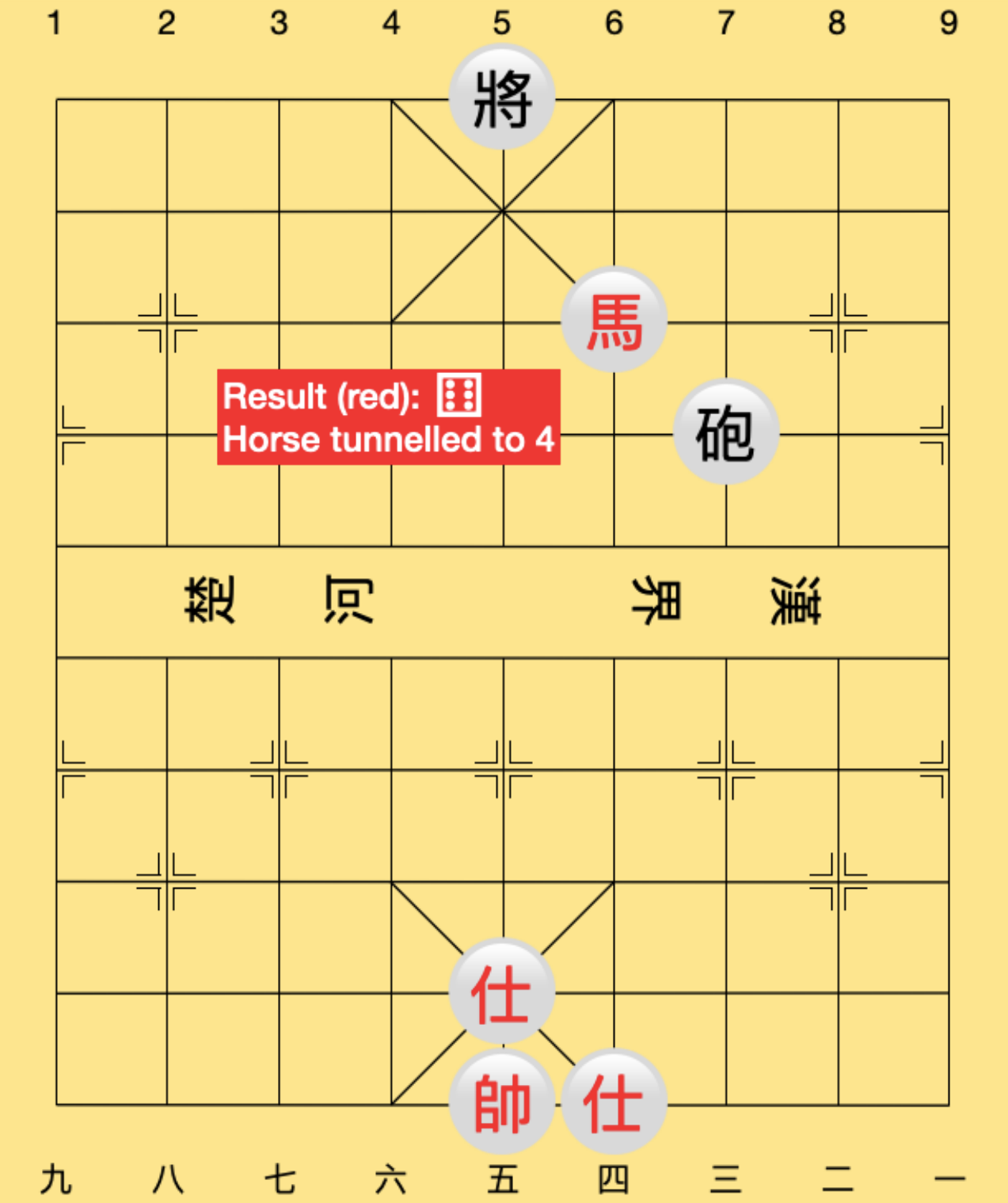}
\end{figure}

\section*{Appendix B: Who is Niel?}\label{niel}

\begin{figure}[H]
\centering
\vspace*{-0.5cm}
\includegraphics[width=0.25\columnwidth]{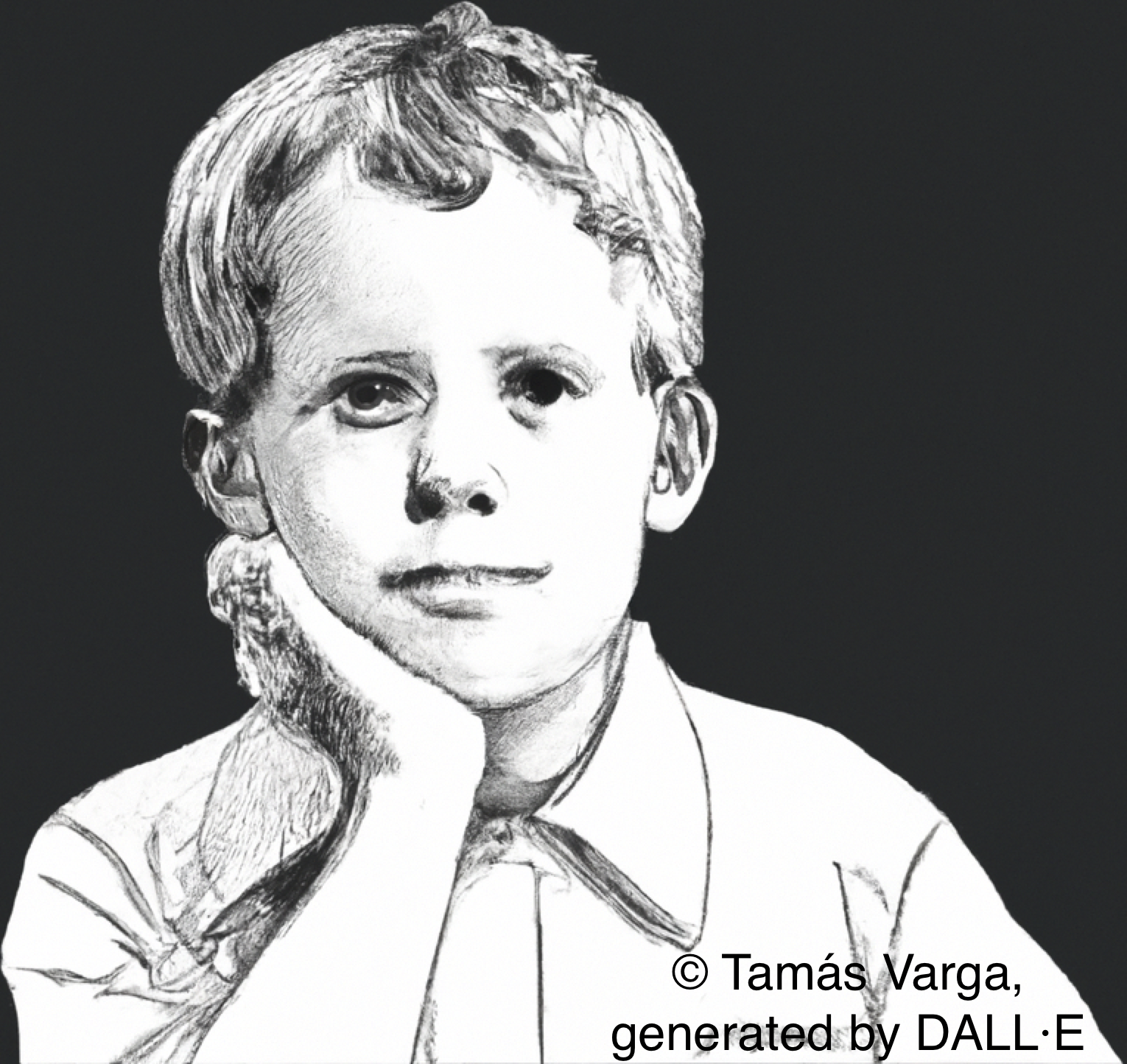}
\end{figure}

\backmatter

\section*{Declarations}

Niel's Chess is a trademark co-owned by the author, and there is a related patent pending in multiple jurisdictions.

\end{document}